\documentclass[final,5p,times,twocolumn]{elsarticle}

\usepackage{amssymb}
\usepackage{color}
\usepackage{amsmath}
\usepackage{multirow}
\usepackage{booktabs}
\usepackage{bm}
\usepackage{braket}
\usepackage{amsmath}
\usepackage{graphicx}
\usepackage{soul}
\usepackage {ulem}
\biboptions{sort&compress}
\usepackage{subcaption}
\usepackage{tabularx}

\usepackage[utf8]{inputenc}  
\usepackage[T1]{fontenc}  

\usepackage[pdfstartview=FitH, colorlinks,
 linkcolor={red!60!black!},
 anchorcolor={green!80!black!},
 urlcolor={blue!80!black!},
 citecolor={blue!50!black!}]{hyperref}
\usepackage[table]{xcolor}

\journal{Physics Letters B}

\newcolumntype{Y}{>{\centering\arraybackslash}X}

\begin{document}


\begin{frontmatter}



\title{\textit{Ab initio} calculations of anomalous seniority breaking in the $\pi g_{9/2}$ shell for the $N=50$ isotones
\tnoteref{label1}}
\tnotetext[label1]{This manuscript has been authored by UT-Battelle, LLC, under contract DE-AC05-00OR22725 with the US Department of Energy (DOE). The US government retains and the publisher, by accepting the article for publication, acknowledges that the US government retains a nonexclusive, paid-up, irrevocable, worldwide license to publish or reproduce the published form of this manuscript, or allow others to do so, for US government purposes. DOE will provide public access to these results of federally sponsored research in accordance with the DOE Public Access Plan (\href{http://energy.gov/downloads/doe-public-access-plan}{http://energy.gov/downloads/doe-public-access-plan}).}


\author[ad1,ad2,ad3]{Q. Yuan}
\author[ad4,ad5]{B. S. Hu\corref{correspondence}}

\address[ad1]{CAS Key Laboratory of High Precision Nuclear Spectroscopy, Institute of Modern Physics,
Chinese Academy of Sciences, Lanzhou 730000, China}
\address[ad2]{School of Nuclear Science and Technology, University of Chinese Academy of Sciences, Beijing 100049, China}
\address[ad3]{Southern Center for Nuclear-Science Theory (SCNT), Institute of Modern Physics, Chinese Academy of Sciences, Huizhou 516000, Guangdong Province, China}

\address[ad4]{National Center for Computational Sciences, Oak Ridge National Laboratory, Oak Ridge, Tennessee 37831, USA}
\address[ad5]{Physics Division, Oak Ridge National Laboratory, Oak Ridge, Tennessee 37831, USA}

\cortext[correspondence]{Corresponding author:\\ hub1@ornl.gov (B. S. Hu)}

\begin{abstract}
We performed \textit{ab initio} valence-space in-medium similarity renormalization group (VS-IMSRG) calculations based on chiral two-nucleon and three-nucleon interactions to investigate the anomalous seniority breaking in the neutron number $N=50$ isotones: $^{92}$Mo, $^{94}$Ru, $^{96}$Pd, and $^{98}$Cd. Our calculations well reproduced the measured low-lying spectra and electromagnetic $E2$ transitions in these nuclei, supporting partial seniority conservation in the first $\pi g_{9/2}$ shell. 
Recent experiments have revealed that, compared to the symmetric patterns predicted under the conserved seniority symmetry, the $4^+_1\to2^+_1$ $E2$ transition strength in $^{94}$Ru is significantly enhanced and that in $^{96}$Pd is suppressed. In contrast, the $6^+_1\to 4^+_1$ and $8^+_1\to6^+_1$ transitions exhibit the opposite trend.
We found that this anomalous asymmetry is sensitive to subtle seniority breaking effects, providing a stringent test for state-of-the-art nucleon-nucleon interactions and nuclear models. We analyzed the anomalous asymmetry using VS-IMSRG calculations across various valence spaces. Our \textit{ab initio} results suggest that core excitations of both proton and neutron across the $Z=50$ shell are ascribed to the observed anomalous seniority breaking in the $N=50$ isotones.
\end{abstract}

\begin{keyword}
\textit{ab initio} \sep chiral nuclear force \sep seniority breaking \sep cross-shell excitations \sep $N=50$ isotones
\end{keyword}

\end{frontmatter}

\section{Introduction}  
In nuclear physics, seniority ($\upsilon_{\rm seniority}$) is the number of nucleons that are not paired to angular momentum zero, in short the number of unpaired neutrons ($\nu$) or protons ($\pi$)~\cite{PhysRev.63.367,RevModPhys.48.191}. This concept offers a simple and regular insight into the complex nuclear many-body system~\cite{QI2017616}.
Seniority may be a good quantum number particularly when a high-$j$ ($j\geq7/2$) single-particle orbital dominates the nuclear structure.
The conservation of seniority exhibits characteristic symmetry rules, such as excitation energies independent of the number of valence nucleons; symmetric parabolic, so called U-shaped, $B(E2)$ patterns for seniority-conserving $E2$ transition probability with seniority change $\Delta \upsilon_{\rm seniority}$=0; inverted U-shaped $B(E2)$ patterns for seniority-changing transitions with $\Delta \upsilon_{\rm seniority}$=2 \cite{casten2001,Ressler2004}.

Semimagic nuclei, in which either the number of protons ($Z$) or neutrons ($N$) corresponds to a magic number, provide an excellent laboratory for investigating seniority symmetry. Examples include the $N = 50$ isotones in the $\pi g_{9/2}$ subshell~\cite{PhysRevLett.129.112501}, the $N = 82$ isotones in the $\pi g_{9/2}$~\cite{PhysRevLett.99.132501,PhysRevLett.111.152501} and the $\pi h_{11/2}$~\cite{PhysRevLett.63.860} subshells, the $N=126$ isotones in the $\pi h_{9/2}$~\cite{PhysRevC.67.054310} and the $\pi h_{11/2}$~\cite{PhysRevC.78.061302} subshells, the nickel isotopes in the $\nu g_{9/2}$ subshell~\cite{PhysRevLett.113.182501,PhysRevLett.116.122502}, the tin isotopes in the $\nu f_{7/2}$ subshell~\cite{PhysRevLett.113.132502}, and the lead isotopes in the $\nu g_{9/2}$ subshell~\cite{PhysRevLett.109.162502}. These nuclei exhibit remarkably similar level schemes and corresponding U-shaped $B(E2)$ patterns, strongly indicating that a high-$j$ orbital is well-separated from others~\cite{Walker_Phys.Scr.95.044004(2020), Walker_Book_2020}, suggesting that seniority is approximately conserved. However, perfectly symmetric U-shaped $B(E2)$ values are expected only if the wavefunctions are absolutely pure, with valence nucleons occupying only a single orbital. In all known cases, asymmetry $B(E2)$ values with respect to the midshell has been observed~\cite{Walker_Phys.Scr.95.044004(2020), Walker_Book_2020,PhysRevC.95.014313,PhysRevC.105.L031304,PhysRevLett.129.112501,PhysRevResearch.6.L022038,PhysRevLett.109.162502,PhysRevLett.111.152501,PhysRevLett.113.132502,PhysRevLett.113.182501,PhysRevLett.116.122502}. An intriguing example is the $\pi g_{9/2}$ shell in the $N=50$ isotones, which has recently attracted particular interest due to the exotic seniority breaking effects~\cite{PhysRevLett.87.172501,PhysRevC.70.044314,PhysRevC.83.014307,PhysRevC.95.014313,PhysRevC.105.L031304,PhysRevLett.129.112501,PhysRevResearch.6.L022038,PhysRevC.98.061303,QI2017616,PhysRevLett.100.052501}, as discussed in the following.

The properties of semimagic $^{94}_{44}$Ru$^{}_{50}$ and $^{96}_{46}$Pd$^{}_{50}$, with their valence nucleons located around the half-filled $\pi g_{9/2}$ orbital, are sensitive to the relative isolation of the $\pi g_{9/2}$ subshell. For these two nuclei, the $E2$ transition strengths with seniority change $\Delta \upsilon_{\rm seniority} =0$ are expected to be symmetric and hindered if the seniority is conserved.
However, for $^{94}$Ru, Mach $et\ al.$ reported a lower limit of $B(E2:4^+_1\to 2^+_1) \ge 46 \ \rm e^2fm^4$~\cite{PhysRevC.95.014313}, and later Das $et\ al.$ reported a value of $B(E2:4^+_1\to 2^+_1) = 103(24) \ \rm e^2fm^4$ for the same transition~\cite{PhysRevC.105.L031304}. These measurements suggest a breaking of seniority in the $N=50$ isotones when compared to the LSSM calculations ~\cite{PhysRevC.95.014313,PhysRevC.105.L031304}. Conversely, P$\rm \Acute{e}$rez-Vidal $et\ al.$ recently reported a value of $B(E2:4^+_1\to 2^+_1) = 38(3) \ \rm e^2fm^4$ for $^{94}$Ru, proposing that seniority is largely conserved in the first $\pi g_{9/2}$ orbital~\cite{PhysRevLett.129.112501}. Regardless of the debate on seniority conservation, an anomalous asymmetry has been observed in the $E2$ transitions, where the $4^+_1\to2^+_1$ $E2$ transition strength of $^{94}$Ru is strongly enhanced and that of $^{96}$Pd is suppressed, while the opposite trend is noted for the $6^+_1\to 4^+_1$ and $8^+_1\to6^+_1$ transitions~\cite{PhysRevC.95.014313,PhysRevLett.129.112501}. These observed anomalous $E2$ transitions pose significant challenges to theoretical models~\cite{PhysRevC.95.014313,PhysRevC.105.L031304,PhysRevLett.129.112501}.

Deviations from good seniority in semimagic nuclei may result from mixing due to a seniority non-conserving interaction, configuration mixing with neighboring states, or core excitation across the shell gap~\cite{PhysRevC.95.014313,PhysRevC.105.L031304}.
To elucidate the structure of the $N=50$ isotones and the exotic partial conservation of seniority or its breaking, large-scale shell model(LSSM) calculations have been employed using various interactions within different model spaces~\cite{PhysRevC.95.014313,PhysRevC.105.L031304,PhysRevLett.129.112501,PhysRevResearch.6.L022038,QI2017616}. Within the proton and neutron $gsd(0g,1d,2s)$ model space, LSSM calculations employing a G-matrix based interaction reproduce the anomalous $E2$ transition trends in the $N=50$ isotones. This suggests that neutron particle-hole excitations across the $N=50$ shell gap are responsible for the breakdown of the seniority scheme in the midshell $N=50$ isotones~\cite{PhysRevC.95.014313}. Alternatively, instead of invoking excitations across the $N=50$ shell gap, LSSM calculations within the $fpg(0f_{5/2},1p,0g_{9/2})$ model space attribute the observed seniority symmetry breaking in $^{94}$Ru to the cross-diagonal components of the interaction, which cause a subtle interference between the wave functions of the $4^+_1(\upsilon_{\rm seniority}=2)$ and $4_2^+(\upsilon_{\rm seniority}=4)$ states~\cite{PhysRevC.105.L031304,QI2017616}. Moreover, LSSM calculations within the $fpg$ model space, based on a realistic effective interaction with fitted single-particle energies, achieve overall good agreement with the experimental data for the $N=50$ isotones and suggest that seniority is largely conserved, although they do not reproduce the anomalous $E2$ transition trends~\cite{PhysRevLett.129.112501}.
The anomalous $E2$ transitions in the $N=50$ isotones provide a critical test ground for modern state-of-the-art nuclear models and nucleon-nucleon interactions.

Recent experimental and theoretical endeavors have shed light on the exotic seniority breaking in the $N=50$ isotones. However, notable discrepancies with experimental results persist, and substantial variations among theoretical models remain. 
Therefore, further experimental and theoretical investigations are essential.
Based on realistic nuclear forces derived from chiral effective field theory ($\chi$EFT)~\cite{E.Epelbaum_RevModPhys.81.1773(2009), R.Machleidt_Phys.Rep.503.1(2011)} and advanced nuclear many-body methods~\cite{hergert2020,BARRETT2013131,RevModPhys.87.1067,Hagen_2014,GIMSRG_PhysRevC.99.061302(2019),doi:10.1146/annurev-nucl-101917-021120,DIMSRG_PhysRevC.105.L061303(2022),B.S.Hu_Nat.Phys.12.186(2016),PhysRevC.100.054313,tichai2024}, nuclear {\it ab initio} calculations have made great progress over the past decades. However, probing seniority symmetry from first principles is still lacking. The \textit{ab initio} valence-space in-medium similarity renormalization group (VS-IMSRG)~\cite{PhysRevC.85.061304,Phys.Rep.621.165, PhysRevLett.118.032502,PhysRevC.107.014302,YUAN2024138331,PhysRevC.109.L041301}, which is formulated in terms of continuous unitary transformation, offers a powerful tool for exploring seniority symmetry. In this work, we employ VS-IMSRG computations with realistic nucleon-nucleon (NN) and three-nucleon (3N) forces derived from $\chi$EFT to investigate the properties of the $N=50$ isotones, focusing on the anomalous $E2$ transitions. 

The letter is organized as follows: First, the VS-IMSRG framework is briefly introduced. Subsequently, we present the systematic calculations of the properties for the $N=50$ isotones, including excitation energies and $E2$ transition strengths. After that, we perform calculations within a variety of valence spaces to analyze the anomalous asymmetry in the $E2$ transitions.  Finally, we conclude with a summary of the present work.

\section{Method}

We start with the intrinsic Hamiltonian of an $A$-body nuclear system, which reads
\begin{equation}
    H=\sum_{i<j}^{A}\left(\frac{(\boldsymbol{p}_{i} - \boldsymbol{p}_{j})^2}{2m A} + V_{i j}^{\mathrm{NN}}\right)+\sum_{i<j<k}^{A} V_{i j k}^{3 \mathrm{N}},
    \label{H_in}
\end{equation}
where $\boldsymbol{p}$ is the nucleon momentum in the laboratory, $m$ is the nucleon mass, $V^{\rm NN}$ and  $V^{\rm 3N}$ represent the two-nucleon (NN) and three-nucleon (3N) interactions, respectively. In this work, we employed the chiral NN + 3N interactions EM1.8/2.0~\cite{PhysRevC.83.031301} and $\Delta$NNLO$_{\rm GO}$ \cite{PhysRevC.102.054301}. EM1.8/2.0 yields accurate ground-state energies and spectra for light, medium and even heavy mass nuclei while underpredicting nuclear radii~\cite{hagen2016b,simonis2017,PhysRevLett.126.022501,PhysRevC.105.014302,hebeler2023}. 
$\Delta$NNLO$_{\rm GO}$ includes explicit $\rm{\Delta}$-isobar degrees of freedom and simultaneously optimizes NN and 3N forces at the N$^2$LO level, especially improving descriptions of nuclear radii.

We then rewrite Hamiltonian~(\ref{H_in}) in terms of normal-ordered operators with respect to the Hartree-Fock reference state,
\begin{equation}
\begin{split}
    H=&E_0+\sum_{ij}f_{ij}:a^{\dagger}_ia_j:+\frac{1}{2!^2}\sum_{ijkl}\Gamma_{ijkl}:a^{\dagger}_ia^{\dagger}_ja_la_k:\\
    &+\frac{1}{3!^2}\sum_{ijklmn}W_{ijklmn}:a^{\dagger}_ia^{\dagger}_ja^{\dagger}_ka_na_ma_l:,
\end{split}
\label{H_norm}
\end{equation}
where $E_0$, $f$, $\Gamma$, and $W$ correspond to the normal-ordered zero-, one-, two-, and three-body terms, respectively. In practical calculations, the residual normal-ordered three-body term $W$ is neglected and the contribution of the 3N force is well captured at the normal-ordered two-body level~\cite{hagen2007a,PhysRevLett.109.052501}.

Next, we decouple Hamiltonian~(\ref{H_norm}) from the large Hilbert space to a small valance space using the VS-IMSRG framework, achieved by solving the flow equation,
\begin{equation}
\frac{dH(s)}{ds}=[\eta(s),H(s)],
\label{FE}
\end{equation}
with an anti-Hermitian generator,
\begin{equation}
\eta(s)\equiv\frac{dU(s)}{ds}U^{\dagger}(s)=-\eta^{\dagger}(s).
\end{equation}
where $U(s)$ is the continuous similarity transformation and $s$ is the flow parameter. Within the Magnus formulation~\cite{PhysRevC.92.034331}, the VS-IMSRG provides an efficient tool to derive the valance-space effective Hamiltonian and operators of other observables,
\begin{equation}
\begin{split}
     H_{\rm eff}=e^{\Omega}He^{-\Omega},\\
    O_{\rm eff}=e^{\Omega}Oe^{-\Omega},   
\end{split}\label{magus}
\end{equation}
with $\Omega=\int\eta(s)ds$.

In this work, we adopted the $gsd$ shell as the valence space for both protons and neutrons. The effective Hamiltonians and $E2$ operators for this valence space were consistently decoupled by VS-IMSRG with ensemble normal ordering (ENO), which can approximately capture 3N forces between valence nucleons~\cite{PhysRevLett.118.032502}. 
In practical calculations, we employed the Magnus formalism, truncating all operators at the two-body level, i.e., VS-IMSRG(2)~\cite{PhysRevC.92.034331}. Subsequently, we diagonalized the valence-space effective Hamiltonians using the large-scale shell model code {\tt KSHELL}~\cite{Phys.Commun.244.372}, including up to four-particle--four-hole ($4p-4h$) excitations across the $N = Z = 50$ major shell, and calculated energies and $E2$ transitions. We confirmed that the results are well converged with the $4p-4h$ truncation.

\begin{figure}[ht]
    \centering
    \includegraphics[width=0.42\paperwidth]{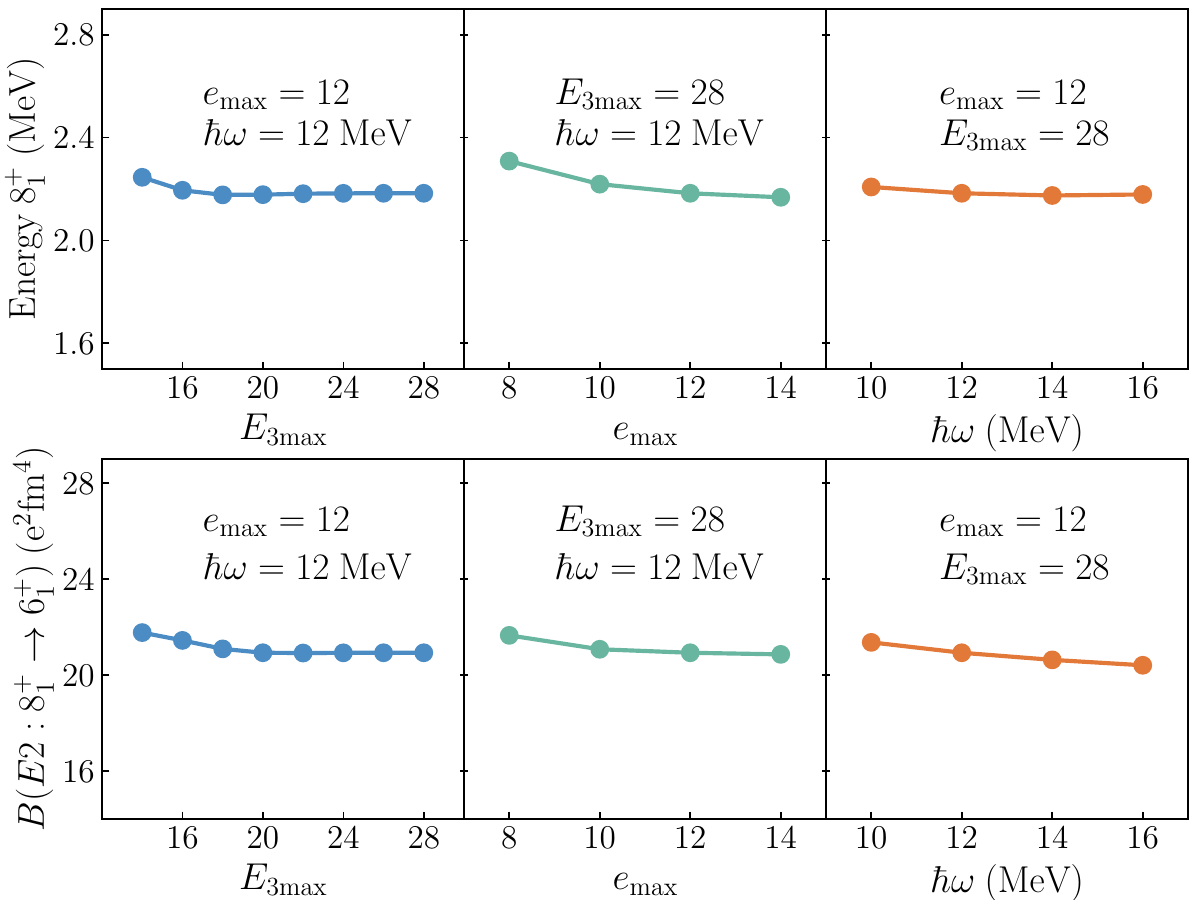}
    \caption{Convergence of the $8^+_1$ energy (top panels) and the $B(E2:8^+_1\to6^+_1)$ value (bottom panels)  for $^{98}$Cd, calculated using VS-IMSRG with the EM1.8/2.0 interaction, as a function of the model space parameters $E_{3\rm max}$, $e_{\rm max}$, and $\hbar\omega$.}
    \label{converge}
\end{figure}

To assess the convergence of the calculations with respect to the chosen model space, we performed VS-IMSRG calculations in different harmonic-oscillator bases, varying the spacing $\hbar\omega$, single-particle energies up to $e_{\rm max}\hbar\omega$, and excitation energies of three nucleons in the 3N interaction up to $E_{\rm 3max}\hbar \omega$.  Fig.~\ref{converge} presents the results of the $8^+_1$ energy (top panels) and the $B(E2:8^+_1\to6^+_1)$ value (bottom panels) for $^{98}$Cd with various $E_{3\rm max}$, $e_{\rm max}$, and $\hbar\omega$. Good convergence for both excitation energies and $B(E2)$ values is achieved within $E_{3\rm max}=$ 28, $e_{\rm max}=$ 12, and $\hbar\omega=$ 12 MeV, which are the model space used in the following sections. The convergence behavior is similar for other states and nuclei investigated in this work.

\section{Results}

\begin{figure*}[!htb]
    \centering
    \includegraphics[width=0.85\paperwidth]{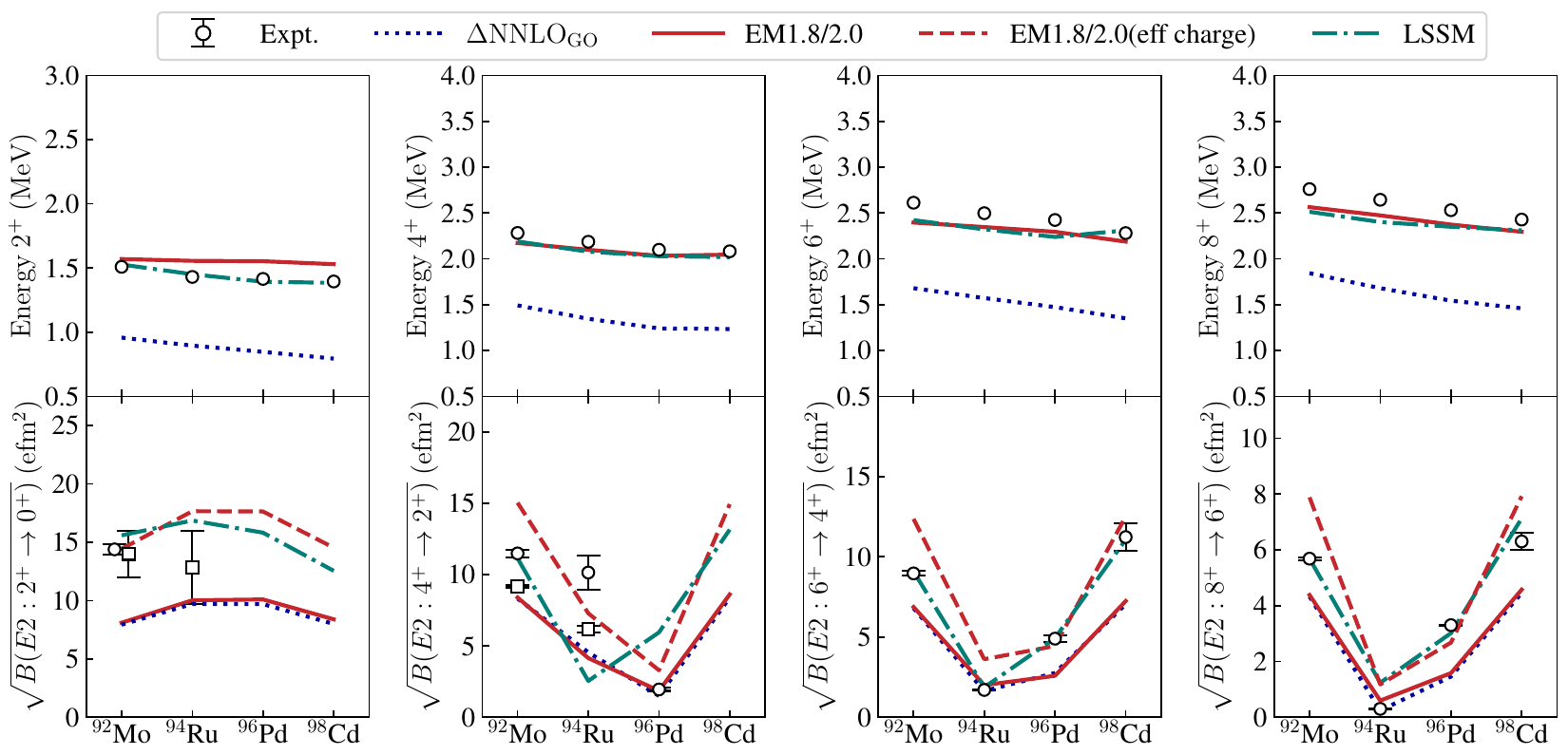}
    \caption{Excitation energies and $B(E2)$ values for the $N=50$ isotones $^{92}$Mo, $^{94}$Ru, $^{96}$Pd, and $^{98}$Cd, calculated by VS-IMSRG within the $gsd$ valence space, compared to LSSM calculations within the $fpg$ model space~\cite{PhysRevLett.129.112501} and available experimental data. For the $E2$ transitions, the blue dotted lines and red solid lines show the VS-IMSRG calculations with consistently transformed $E2$ operators using the $\Delta\rm NNLO_{GO}$ and EM1.8/2.0 interactions, respectively. The red dashed lines represent the VS-IMSRG calculations using the standard empirical effective charges $e_\pi=1.5,\ e_\nu=0.5$, and the green dot-dashed lines indicate the LSSM calculations using the microscopic effective charges~\cite{PhysRevLett.129.112501}, respectively. The open squares represent the experimental data taken from Ref.~\cite{PhysRevLett.129.112501} and the open circles are the experimental data taken from Refs.~\cite{nndc,PhysRevC.95.014313,PhysRevC.96.044311,PhysRevC.103.049901,PhysRevC.105.L031304,PhysRevC.108.064313}.}
    \label{N50_Ex_BE2}
\end{figure*}

\begin{figure}
    \centering
    \includegraphics[width=0.35\paperwidth]{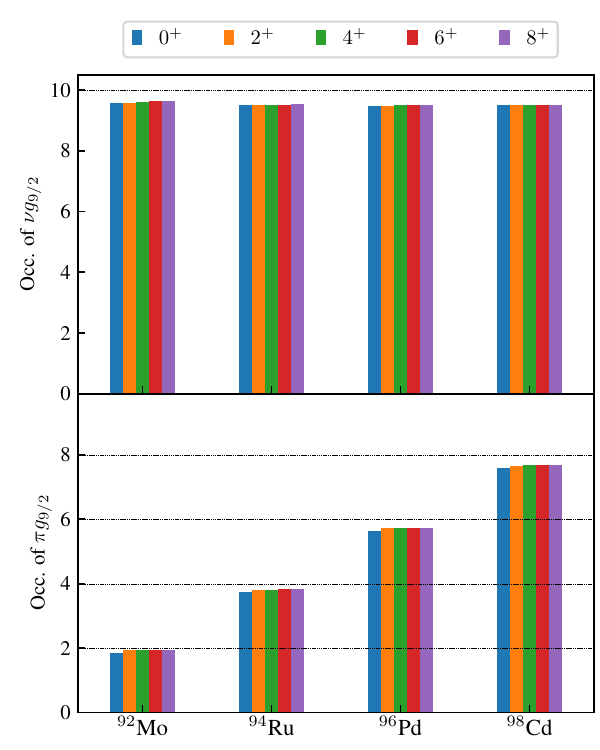}
    \caption{Occupation numbers in the neutron $\nu g_{9/2}$ orbit and the proton $\pi g_{9/2}$ orbit for the $N=50$ isotones $^{92}$Mo, $^{94}$Ru, $^{96}$Pd, and $^{98}$Cd,  calculated by VS-IMSRG within the $gsd$ valence space using the EM1.8/2.0 interaction.}
    \label{Occ}
\end{figure}

As discussed in the Introduction section, the anomalous $E2$ transitions observed in the $N=50$ isotones, indicative of some degree of seniority breaking, pose significant challenges to theoretical models~\cite{PhysRevC.95.014313,PhysRevC.105.L031304,PhysRevLett.129.112501}. It is anticipated that \textit{ab initio} calculations, using high-resolution interactions derived from the effective field theory of quantum chromodynamics, will provide a comprehensive and robust understanding of these anomalies.
In this letter, we systematically calculated the spectra and $E2$ transitions for the $N=50$ isotones $^{92}$Mo, $^{94}$Ru, $^{96}$Pd, and $^{98}$Cd from first principles. Fig.~\ref{N50_Ex_BE2} presents our \textit{ab initio} results for the excitation energies of the $2^+$, $4^+$, $6^+$, and $8^+$ yrast states, along with the corresponding $B(E2)$ values for the transitions from $2^+_1 \to 0^+_1$, $4^+_1 \to 2^+_1$, $6^+_1 \to 4^+_1$, and $8^+_1 \to 6^+_1$. These results, obtained using the VS-IMSRG approach within the $gsd$ valence space, are compared with the LSSM calculations within the $fpg$ valence space~\cite{PhysRevLett.129.112501} and recent experimental data~\cite{PhysRevLett.129.112501,nndc,PhysRevC.95.014313,PhysRevC.96.044311,PhysRevC.103.049901,PhysRevC.105.L031304,PhysRevC.108.064313}.

The upper panel of Fig.~\ref{N50_Ex_BE2} shows that the \textit{ab initio} yrast level schemes calculated using EM1.8/2.0 interaction closely align with the $fpg$-shell LSSM calculations~\cite{PhysRevLett.129.112501} and accurately reproduce the experimental data. The yrast levels remain nearly constant across the $N=50$ isotones, demonstrating the characteristic of approximate seniority conservation. Fig.~\ref{Occ} illustrates occupation numbers in the $\nu_{g_{9/2}}$ and $\pi_{g_{9/2}}$ orbits, indicating that the $|(\pi g_{9/2})^n \rangle$ configuration of $n$ valence nucleons predominantly determines the structure of the $N=50$ isotones, implying that the seniority is largely conserved. Additionally, our results reveal significant energy disparities between the $4^+_1$ and $4^+_2$ states in $^{92}$Mo and $^{98}$Cd, with the $4^+_2$ state being more than 3.5 MeV higher in energy than the $4^+_1$ state. In contrast, in $^{94}$Ru and $^{96}$Pd, the $4^+_2$ and $4^+_1$ states are very close in energy, separated by less than 500 keV. Such proximity may facilitate seniority mixing or, in other words, result in interference between the $4^+_2$ and $4^+_1$ states in $^{94}$Ru and $^{96}$Pd, as discussed in previous studies~\cite{PhysRevC.105.L031304,QI2017616}. VS-IMSRG spectra from the $\Delta\rm NNLO_{GO}$ interaction exhibit a nearly constant shift relative to the EM1.8/2.0 results. We find that by replacing the $\langle \pi g_{9/2}, \pi g_{9/2}; J=0 | H_{\rm VS-IMSRG} | \pi g_{9/2}, \pi g_{9/2}; J=0 \rangle$ valence-space matrix element with the one from EM1.8/2.0, the adjusted $\Delta\rm NNLO_{GO}$ excitation energies align closely with those of EM1.8/2.0. The matrix element actually represents the dominant pairing interaction. Therefore, variations in the partial wave $^1S_0$ of nucleon-nucleon force, which plays a central role in setting up pairing correlations in nuclei \cite{sun2024}, can account for the level differences between $\Delta\rm NNLO_{GO}$ and EM1.8/2.0. 


We turn to the reduced transition strengths $B(E2)$. If the seniority symmetry is conserved, the $E2$ transition matrix elements for configuration $|j^n, J \rangle$, which consists of $n$ valence nucleons in a single-$j$ shell with total angular momentum $J$, are well known \cite{casten2001,Ressler2004},
\begin{equation}
\begin{split}
&\langle j^n \upsilon_{\rm seniority}, J_{f} || E2 || j^n \upsilon_{\rm seniority}, J_{i} \rangle
= \sqrt{B(E2:i \rightarrow f)\cdot (2J_i+1)} \\ &
= \frac{2j+1-2n}{2j+1-2\upsilon_{\rm seniority}} \langle j^{\upsilon_{\rm seniority}} \upsilon_{\rm seniority}, J_{f} || E2 || j^{\upsilon_{\rm seniority}} \upsilon_{\rm seniority}, J_{i} \rangle,
\end{split}
\label{dv0}
\end{equation}
\begin{equation}
\begin{split}
&\langle j^n \upsilon_{\rm seniority}, J_{f} || E2 || j^n \upsilon_{\rm seniority}-2, J_{i} \rangle  \\ &
= \sqrt{B(E2:i \rightarrow f)\cdot (2J_i+1)} \\ &
= \left[ \frac{ (n-\upsilon_{\rm seniority}+2)(2j+3-n-\upsilon_{\rm seniority})}{2(2j+3-2\upsilon_{\rm seniority})} \right]^{1/2} \\ &
\cdot \langle j^{\upsilon_{\rm seniority}} \upsilon_{\rm seniority}, J_{f} || E2 || j^{\upsilon_{\rm seniority}} \upsilon_{\rm seniority}-2, J_{i} \rangle.
\end{split}
\label{dv2}
\end{equation}
According to Eq.~(\ref{dv0}), for seniority-conserving ($\Delta \upsilon_{\rm seniority} = 0$) $E2$ transitions with fixed $j$ and $ \upsilon_{\rm seniority}$ across different $n$ or nuclei, the configuration $|j^{\upsilon_{\rm seniority}}, J \rangle$  is independent of $n$. Thus, $B(E2) \propto \left(\dfrac{2j+1}{2} -n \right)^2$. In this scenario, the $B(E2)$ values for transitions  such as $4^+_1\to 2^+_1$, $6^+_1\to 4^+_1$, and $8^+_1\to 6^+_1$ exhibit a symmetric U-shape as a function of $n$, minimizing at mid-$j$ subshell, where $n=(2j+1)/2$. Conversely, for the seniority-nonconserving ($\Delta \upsilon_{\rm seniority} = 2$) $E2$ transitions, such as $2^+_1\to 0^+_1$, $B(E2) \propto (2-\upsilon_{\rm seniority}+n)(2j+3-\upsilon_{\rm seniority}-n)$, displaying an inverted U-shape as a function of $n$ and peaking at the midshell, $n=(2j+1)/2$, as described by Eq.~(\ref{dv2}).
Recent observations of the $E2$ transition strengths in the $N=50$ isotones generally align with these expectations, suggesting partial conservation of seniority~\cite{PhysRevLett.129.112501}. However, anomalies are evident: the $4^+_1\to2^+_1$ $E2$ transition strength in $^{94}$Ru is unexpectedly high, while it is diminished in $^{96}$Pd, hinting at some degree of seniority breaking~\cite{PhysRevC.95.014313,PhysRevLett.129.112501}. Morevoer, the trend observed in the $4^+_1\to2^+_1$ $E2$ transition across $N=50$ isotones contrasts with the tread in the $6^+_1\to 4^+_1$ and $8^+_1\to6^+_1$ transitions, where $B(E2)$ values for $^{94}$Ru are lower than those for $^{96}$Pd~\cite{PhysRevC.95.014313,PhysRevLett.129.112501}.

\begin{figure*}[!htb]
    \centering
    \includegraphics[width=0.7\paperwidth]{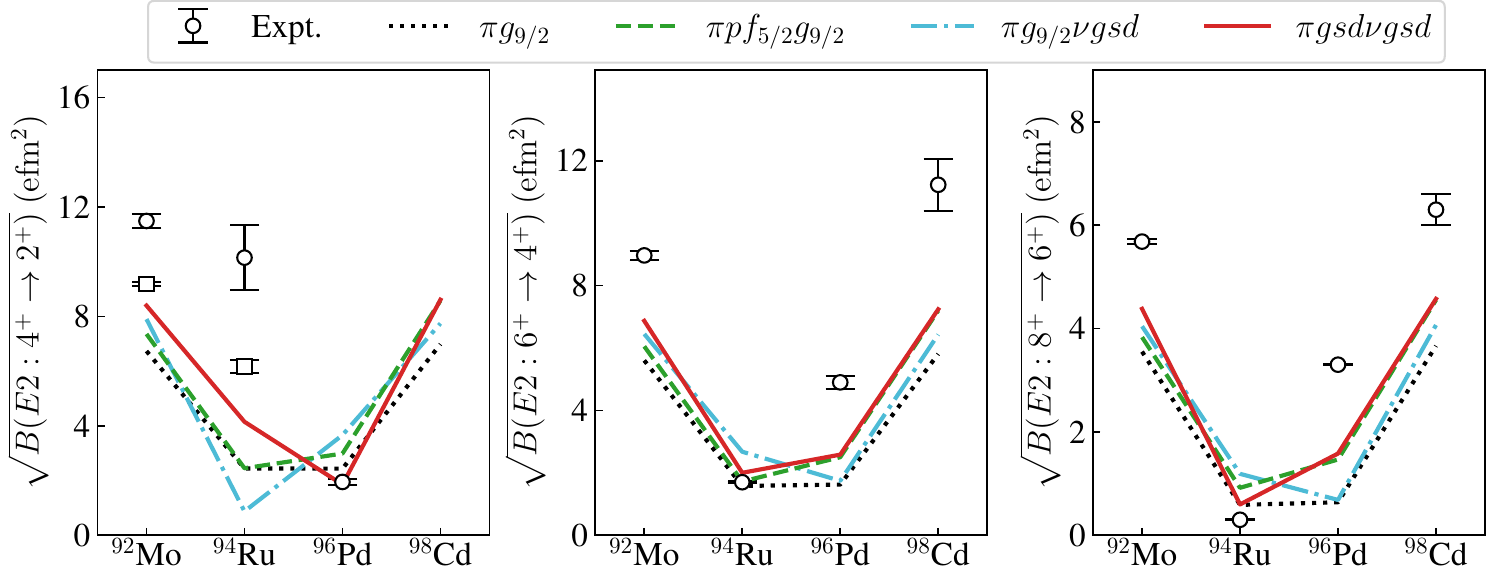}
    \caption{$B(E2)$ values for $^{92}$Mo, $^{94}$Ru, $^{96}$Pd, and $^{98}$Cd, calculated by VS-IMSRG within a variety of valence spaces. The open squares represent the experimental data taken from Ref.~\cite{PhysRevLett.129.112501} and the open circles label the experimental data taken from Refs.~\cite{nndc,PhysRevC.95.014313,PhysRevC.96.044311,PhysRevC.103.049901,PhysRevC.105.L031304,PhysRevC.108.064313}.}
    \label{N50_BE2_468}
\end{figure*}

For $E2$ transition strengths, LSSM calculations using various effective interactions, model spaces, and empirical effective charges have shown significant discrepancies~\cite{PhysRevC.95.014313,PhysRevC.105.L031304,PhysRevLett.129.112501,PhysRevResearch.6.L022038,QI2017616}. 
Fig.~\ref{N50_Ex_BE2} displays one of the best LSSM results, which achieves good overall agreement with the experimental data within the $fpg$ model space~\cite{PhysRevLett.129.112501}. However, it fails to reproduce the anomalous $E2$ transition trends observed in $^{94}$Ru and $^{96}$Pd.
While the LSSM calculations within the $gsd$ valence space from Ref.~\cite{PhysRevC.95.014313} can reproduce the observed $E2$ transition trends, they provide a poor description of the experimental $B(E2)$ values. This suggests that the anomalous behavior may stem from a breakdown of the seniority quantum number within the $\pi g_{9/2}$ configuration, driven by neutron particle-hole excitations across the $N = 50$ shell. Further investigation into the causes of these anomalous $B(E2)$ behaviors is needed.

As depicted in Fig.~\ref{N50_Ex_BE2}, our VS-IMSRG results, employing the consistently derived $gsd$-valence-space Hamiltonian and $E2$ operator, successfully capture the observed anomalous $E2$ transition trends and provide a good description of overall $B(E2)$ values.
We also present results calculated using the standard empirical effective charges $e_\pi=1.5$ and $e_\nu=0.5$ rather than the consistently evolved $E2$ operator, showing better alignment with recent experimental data. 
Although the VS-IMSRG(2) often underestimates the absolute value of quadrupole moments and $B(E2)$, primarily due to the absence of higher-order collective excitations \cite{Hend18E2,BAI2022137064}, it provides a reasonable description of the $B(E2)$ trends for the $N=50$ isotones, which are characterized by largely conserved seniority.
When comparing $B(E2)$ calculations using EM1.8/2.0 and $\Delta\rm NNLO_{GO}$ interactions, we find that both interactions yield nearly identical results. Although the EM1.8/2.0 interaction underestimates nuclear radii \cite{hagen2016b,simonis2017}, it does not significantly underpredict $B(E2)$ values as previous findings~\cite{sun2024,hu2024}.

To understand the observed anomalous $E2$ transitions, we conducted VS-IMSRG calculations within various valence spaces using consistently evolved $E2$ operators. Although the yrast spectra calculated by VS-IMSRG across different valence spaces are generally quite similar, the calculated $E2$ transition strengths reveal distinct patterns, as shown in Fig.~\ref{N50_BE2_468}. The black dotted lines in Fig.~\ref{N50_BE2_468} represent results calculated within the valence space confined to the $\pi g_{9/2}$ orbital only. 
In this calculation, seniority is conserved, resulting in perfectly symmetric U-shaped $B(E2)$ values for the $4^+_1 \to 2^+_1$, $6^+_1 \to 4^+_1$, and $8^+_1 \to 6^+_1$ transitions.
When including excitations from the proton $fp$ shell, depicted by the green dashed lines in Fig.~\ref{N50_BE2_468}, the VS-IMSRG calculations within the $\pi{pf_{5/2}g_{9/2}}$ valence space reproduce most $E2$ transition trends across the $N=50$ isotonic chain, except for the $4^+_1 \to 2^+_1$ $E2$ transition. These $B(E2)$ trends generally align with the results from the LSSM calculations in the $fpg$ valence space~\cite{PhysRevLett.129.112501}.
Within the $fpg$ valence space~\cite{PhysRevLett.129.112501}, particle-hole excitations across the $Z=50$ and $N=50$ shells can not be included. However, cross-shell excitations across the $^{100}$Sn core have been observed in the $N=50$ nuclei below $^{100}$Sn~\cite{PhysRevC.69.064304,JUNGCLAUS1998346}. The deviations between our calculations within the $\pi{pf_{5/2}g_{9/2}}$ valence space and experimental observations for the $4^+_1 \to 2^+_1$ transition in $^{94}$Ru and $^{96}$Pd indicate that excitations across the $Z=50$ and $N=50$ shell gaps may play a significant role.

We extended VS-IMSRG calculations to include excitations across the $Z=50$ and $N=50$ shell gaps. We first performed calculations within the $\pi{g_{9/2}}\nu{gsd}$ valence space, which incorporates neutron excitations across the $N=50$ shell. Despite observing seniority breaking effects compared to $\pi g_{9/2}$-only results, the trends for the $4^+_1 \to 2^+_1$, $6^+_1 \to 4^+_1$, and $8^+_1 \to 6^+_1$ $E2$ transitions are not consistent with the experimental data, as illustrated by the cyan dot-dashed lines in Fig.~\ref{N50_BE2_468}. 
This indicates that neutron particle-hole excitations across the $N=50$ shell result in seniority breaking but do not explain the observed anomalous $E2$ transitions.
Finally, we conducted calculations within the $\pi{gsd}\nu{gsd}$ valence space, further including proton excitations across the $Z=50$ shell. As shown by the red solid lines in Fig.~\ref{N50_BE2_468}, the results that included both proton and neutron excitations across the $^{100}$Sn core successfully reproduce all observed $E2$ transition trends. We thus conclude that proton core excitations across the $Z=50$ shell are responsible for the anomalous behavior in the $E2$ transitions of the $N=50$ isotones. Note that significant discrepancies exist among different experimental results~\cite{nndc,PhysRevC.95.014313,PhysRevC.96.044311,PhysRevC.103.049901,PhysRevC.105.L031304,PhysRevC.108.064313}, as shown in Fig.~\ref{N50_BE2_468}. These experimental $B(E2)$ values are inferred using various methods and may remain somewhat model-dependent.
The present work highlights the need for further experimental studies that can significantly reduce model dependence, such as those using Coulomb excitation, 
to better understand the role of seniority breaking in this isotonic chain.

\section{Summary}

Starting with the chiral two-nucleon and three-nucleon interactions, we presented \textit{ab initio} VS-IMSRG calculations of spectra and $E2$ transition strengths for the $N=50$ isotones, focusing on the investigation of the exotic seniority breaking in the first $\pi g_{9/2}$ shell. To understand the observed $E2$ transition properties, we performed VS-IMSRG computations across various valence spaces. Overall, our calculations, which included proton excitations across the $^{100}$Sn core, align well with recent experimental data, particularly in reproducing the observed anomalous $E2$ transition trends. We concluded that both proton and neutron core excitations across the $Z=50$ shell predominantly drive the anomalous asymmetric behavior observed in the $E2$ transitions in the $N=50$ isotones.

The present work demonstrates the capability of \textit{ab initio} calculations, using chiral NN and 3N forces derived from $\chi$EFT, to robustly address seniority symmetry and magicity. Moreover, it motivates more comprehensive experimental investigations. Work is currently in progress to probe observed seniority breaking effects in semimagic nuclei across other regions of the nuclear chart.
 
\section*{Acknowledgments}

We thank Takayuki Miyagi for the {\tt NuHamil} code~\cite{miyagi2023} used to generate matrix elements of the chiral three-body interaction, and Ragnar Stroberg for the {\tt imsrg++} code~\cite{Stro17imsrg++} used to perform VS-IMSRG decoupling.
This work has been supported by the National Key R\&D Program of China under Grant No. 2023YFA1606403; the National Natural Science Foundation of China under Grant Nos.  12205340, 12347106, and 12121005;  the Gansu Natural Science Foundation under Grant No. 22JR5RA123 and 23JRRA614; the Key Research Program of the Chinese Academy of Sciences under Grant No. XDPB15; the State Key Laboratory of Nuclear Physics and Technology, Peking University under Grant No. NPT2020KFY13; and the U.S. Department of Energy (DOE), Office of
Science, under SciDAC-5 (NUCLEI collaboration) and  Contract No. DE-AC05-00OR22725. The numerical calculations in this paper have been done on Hefei advanced computing center.

\section*{References}

\bibliographystyle{elsarticle-num-names}
\bibliography{Ref.bib}

\begin{thebibliography}{66}
\expandafter\ifx\csname natexlab\endcsname\relax\def\natexlab#1{#1}\fi
\providecommand{\url}[1]{\texttt{#1}}
\providecommand{\href}[2]{#2}
\providecommand{\path}[1]{#1}
\providecommand{\DOIprefix}{doi:}
\providecommand{\ArXivprefix}{arXiv:}
\providecommand{\URLprefix}{URL: }
\providecommand{\Pubmedprefix}{pmid:}
\providecommand{\doi}[1]{\href{http://dx.doi.org/#1}{\path{#1}}}
\providecommand{\Pubmed}[1]{\href{pmid:#1}{\path{#1}}}
\providecommand{\bibinfo}[2]{#2}
\ifx\xfnm\relax \def\xfnm[#1]{\unskip,\space#1}\fi
\bibitem[{Racah(1943)}]{PhysRev.63.367}
\bibinfo{author}{G.~Racah},
\newblock \bibinfo{title}{Theory of complex spectra. iii},
\newblock \bibinfo{journal}{Phys. Rev.} \bibinfo{volume}{63}
  (\bibinfo{year}{1943}) \bibinfo{pages}{367--382}. \URLprefix
  \url{https://link.aps.org/doi/10.1103/PhysRev.63.367}.
  \DOIprefix\doi{10.1103/PhysRev.63.367}.
\bibitem[{Schiffer and True(1976)}]{RevModPhys.48.191}
\bibinfo{author}{J.~P. Schiffer}, \bibinfo{author}{W.~W. True},
\newblock \bibinfo{title}{The effective interaction between nucleons deduced
  from nuclear spectra},
\newblock \bibinfo{journal}{Rev. Mod. Phys.} \bibinfo{volume}{48}
  (\bibinfo{year}{1976}) \bibinfo{pages}{191--217}. \URLprefix
  \url{https://link.aps.org/doi/10.1103/RevModPhys.48.191}.
  \DOIprefix\doi{10.1103/RevModPhys.48.191}.
\bibitem[{Qi(2017)}]{QI2017616}
\bibinfo{author}{C.~Qi},
\newblock \bibinfo{title}{Partial conservation of seniority and its unexpected
  influence on {E2} transitions in g9/2 nuclei},
\newblock \bibinfo{journal}{Phys. Lett. B} \bibinfo{volume}{773}
  (\bibinfo{year}{2017}) \bibinfo{pages}{616--619}. \URLprefix
  \url{https://www.sciencedirect.com/science/article/pii/S0370269317307232}.
  \DOIprefix\doi{https://doi.org/10.1016/j.physletb.2017.09.025}.
\bibitem[{Casten(2001)}]{casten2001}
\bibinfo{author}{R.~F. Casten}, \bibinfo{title}{Nuclear {Structure} from a
  {Simple} {Perspective}}, \bibinfo{publisher}{Oxford University Press},
  \bibinfo{year}{2001}. \URLprefix
  \url{https://doi.org/10.1093/acprof:oso/9780198507246.001.0001}.
  \DOIprefix\doi{10.1093/acprof:oso/9780198507246.001.0001}.
\bibitem[{Ressler et~al.(2004)Ressler, Casten, Zamfir, Beausang, Cakirli, Ai,
  Amro, Caprio, Hecht, Heinz, Langdown, McCutchan, Meyer, Plettner, Regan,
  Sciacchitano, and Yamamoto}]{Ressler2004}
\bibinfo{author}{J.~J. Ressler}, \bibinfo{author}{R.~F. Casten},
  \bibinfo{author}{N.~V. Zamfir}, \bibinfo{author}{C.~W. Beausang},
  \bibinfo{author}{R.~B. Cakirli}, \bibinfo{author}{H.~Ai},
  \bibinfo{author}{H.~Amro}, \bibinfo{author}{M.~A. Caprio},
  \bibinfo{author}{A.~A. Hecht}, \bibinfo{author}{A.~Heinz},
  \bibinfo{author}{S.~D. Langdown}, \bibinfo{author}{E.~A. McCutchan},
  \bibinfo{author}{D.~A. Meyer}, \bibinfo{author}{C.~Plettner},
  \bibinfo{author}{P.~H. Regan}, \bibinfo{author}{M.~J.~S. Sciacchitano},
  \bibinfo{author}{A.~D. Yamamoto},
\newblock \bibinfo{title}{Transition from the seniority regime to collective
  motion},
\newblock \bibinfo{journal}{Phys. Rev. C} \bibinfo{volume}{69}
  (\bibinfo{year}{2004}) \bibinfo{pages}{034317}. \URLprefix
  \url{https://link.aps.org/doi/10.1103/PhysRevC.69.034317}.
  \DOIprefix\doi{10.1103/PhysRevC.69.034317}.
\bibitem[{P\'erez-Vidal et~al.(2022)P\'erez-Vidal, Gadea, Domingo-Pardo,
  Gargano, Valiente-Dob\'on, Cl\'ement, Lemasson, Coraggio, Siciliano, Szilner,
  Bast, Braunroth, Collado, Corina, Dewald, Doncel, Dudouet, de~France,
  Fransen, Gonz\'alez, H\"uy\"uk, Jacquot, John, Jungclaus, Kim, Korichi,
  Labiche, Lenzi, Li, Ljungvall, L\'opez-Martens, Mengoni, Michelagnoli,
  M\"uller-Gatermann, Napoli, Navin, Quintana, Ramos, Rejmund, Sanchis,
  Simpson, Stezowski, Wilmsen, Zieli\ifmmode~\acute{n}\else \'{n}\fi{}ska,
  Boston, Barrientos, Bednarczyk, Benzoni, Birkenbach, Boston, Bracco,
  Cederwall, Cullen, Didierjean, Eberth, Gottardo, Goupil, Harkness-Brennan,
  Hess, Judson, Ka\ifmmode \mbox{\c{s}}\else
  \c{s}\fi{}ka\ifmmode~\mbox{\c{s}}\else \c{s}\fi{}, Korten, Leoni, Menegazzo,
  Million, Nyberg, Podolyak, Pullia, Ralet, Recchia, Reiter, Rezynkina, Salsac,
  \ifmmode \mbox{\c{S}}\else \c{S}\fi{}enyi\ifmmode~\breve{g}\else
  \u{g}\fi{}it, Sohler, Theisen, and Verney}]{PhysRevLett.129.112501}
\bibinfo{author}{R.~M. P\'erez-Vidal}, \bibinfo{author}{A.~Gadea},
  \bibinfo{author}{C.~Domingo-Pardo}, \bibinfo{author}{A.~Gargano},
  \bibinfo{author}{J.~J. Valiente-Dob\'on}, \bibinfo{author}{E.~Cl\'ement},
  \bibinfo{author}{A.~Lemasson}, \bibinfo{author}{L.~Coraggio},
  \bibinfo{author}{M.~Siciliano}, \bibinfo{author}{S.~Szilner},
  \bibinfo{author}{M.~Bast}, \bibinfo{author}{T.~Braunroth},
  \bibinfo{author}{J.~Collado}, \bibinfo{author}{A.~Corina},
  \bibinfo{author}{A.~Dewald}, \bibinfo{author}{M.~Doncel},
  \bibinfo{author}{J.~Dudouet}, \bibinfo{author}{G.~de~France},
  \bibinfo{author}{C.~Fransen}, \bibinfo{author}{V.~Gonz\'alez},
  \bibinfo{author}{T.~H\"uy\"uk}, \bibinfo{author}{B.~Jacquot},
  \bibinfo{author}{P.~R. John}, \bibinfo{author}{A.~Jungclaus},
  \bibinfo{author}{Y.~H. Kim}, \bibinfo{author}{A.~Korichi},
  \bibinfo{author}{M.~Labiche}, \bibinfo{author}{S.~Lenzi},
  \bibinfo{author}{H.~Li}, \bibinfo{author}{J.~Ljungvall},
  \bibinfo{author}{A.~L\'opez-Martens}, \bibinfo{author}{D.~Mengoni},
  \bibinfo{author}{C.~Michelagnoli}, \bibinfo{author}{C.~M\"uller-Gatermann},
  \bibinfo{author}{D.~R. Napoli}, \bibinfo{author}{A.~Navin},
  \bibinfo{author}{B.~Quintana}, \bibinfo{author}{D.~Ramos},
  \bibinfo{author}{M.~Rejmund}, \bibinfo{author}{E.~Sanchis},
  \bibinfo{author}{J.~Simpson}, \bibinfo{author}{O.~Stezowski},
  \bibinfo{author}{D.~Wilmsen},
  \bibinfo{author}{M.~Zieli\ifmmode~\acute{n}\else \'{n}\fi{}ska},
  \bibinfo{author}{A.~J. Boston}, \bibinfo{author}{D.~Barrientos},
  \bibinfo{author}{P.~Bednarczyk}, \bibinfo{author}{G.~Benzoni},
  \bibinfo{author}{B.~Birkenbach}, \bibinfo{author}{H.~C. Boston},
  \bibinfo{author}{A.~Bracco}, \bibinfo{author}{B.~Cederwall},
  \bibinfo{author}{D.~M. Cullen}, \bibinfo{author}{F.~Didierjean},
  \bibinfo{author}{J.~Eberth}, \bibinfo{author}{A.~Gottardo},
  \bibinfo{author}{J.~Goupil}, \bibinfo{author}{L.~J. Harkness-Brennan},
  \bibinfo{author}{H.~Hess}, \bibinfo{author}{D.~S. Judson},
  \bibinfo{author}{A.~Ka\ifmmode \mbox{\c{s}}\else
  \c{s}\fi{}ka\ifmmode~\mbox{\c{s}}\else \c{s}\fi{}},
  \bibinfo{author}{W.~Korten}, \bibinfo{author}{S.~Leoni},
  \bibinfo{author}{R.~Menegazzo}, \bibinfo{author}{B.~Million},
  \bibinfo{author}{J.~Nyberg}, \bibinfo{author}{Z.~Podolyak},
  \bibinfo{author}{A.~Pullia}, \bibinfo{author}{D.~Ralet},
  \bibinfo{author}{F.~Recchia}, \bibinfo{author}{P.~Reiter},
  \bibinfo{author}{K.~Rezynkina}, \bibinfo{author}{M.~D. Salsac},
  \bibinfo{author}{M.~\ifmmode \mbox{\c{S}}\else
  \c{S}\fi{}enyi\ifmmode~\breve{g}\else \u{g}\fi{}it},
  \bibinfo{author}{D.~Sohler}, \bibinfo{author}{C.~Theisen},
  \bibinfo{author}{D.~Verney},
\newblock \bibinfo{title}{Evidence of partial seniority conservation in the
  $\ensuremath{\pi}{g}_{9/2}$ shell for the ${N}=50$ isotones},
\newblock \bibinfo{journal}{Phys. Rev. Lett.} \bibinfo{volume}{129}
  (\bibinfo{year}{2022}) \bibinfo{pages}{112501}. \URLprefix
  \url{https://link.aps.org/doi/10.1103/PhysRevLett.129.112501}.
  \DOIprefix\doi{10.1103/PhysRevLett.129.112501}.
\bibitem[{Jungclaus et~al.(2007)Jungclaus, C\'aceres, G\'orska, Pf\"utzner,
  Pietri, Werner-Malento, Grawe, Langanke, Mart\'{\i}nez-Pinedo, Nowacki,
  Poves, Cuenca-Garc\'{\i}a, Rudolph, Podolyak, Regan, Detistov, Lalkovski,
  Modamio, Walker, Bednarczyk, Doornenbal, Geissel, Gerl, Grebosz, Kojouharov,
  Kurz, Prokopowicz, Schaffner, Wollersheim, Andgren, Benlliure, Benzoni,
  Bruce, Casarejos, Cederwall, Crespi, Hadinia, Hellstr\"om, Hoischen, Ilie,
  Jolie, Khaplanov, Kmiecik, Kumar, Maj, Mandal, Montes, Myalski, Simpson,
  Steer, Tashenov, and Wieland}]{PhysRevLett.99.132501}
\bibinfo{author}{A.~Jungclaus}, \bibinfo{author}{L.~C\'aceres},
  \bibinfo{author}{M.~G\'orska}, \bibinfo{author}{M.~Pf\"utzner},
  \bibinfo{author}{S.~Pietri}, \bibinfo{author}{E.~Werner-Malento},
  \bibinfo{author}{H.~Grawe}, \bibinfo{author}{K.~Langanke},
  \bibinfo{author}{G.~Mart\'{\i}nez-Pinedo}, \bibinfo{author}{F.~Nowacki},
  \bibinfo{author}{A.~Poves}, \bibinfo{author}{J.~J. Cuenca-Garc\'{\i}a},
  \bibinfo{author}{D.~Rudolph}, \bibinfo{author}{Z.~Podolyak},
  \bibinfo{author}{P.~H. Regan}, \bibinfo{author}{P.~Detistov},
  \bibinfo{author}{S.~Lalkovski}, \bibinfo{author}{V.~Modamio},
  \bibinfo{author}{J.~Walker}, \bibinfo{author}{P.~Bednarczyk},
  \bibinfo{author}{P.~Doornenbal}, \bibinfo{author}{H.~Geissel},
  \bibinfo{author}{J.~Gerl}, \bibinfo{author}{J.~Grebosz},
  \bibinfo{author}{I.~Kojouharov}, \bibinfo{author}{N.~Kurz},
  \bibinfo{author}{W.~Prokopowicz}, \bibinfo{author}{H.~Schaffner},
  \bibinfo{author}{H.~J. Wollersheim}, \bibinfo{author}{K.~Andgren},
  \bibinfo{author}{J.~Benlliure}, \bibinfo{author}{G.~Benzoni},
  \bibinfo{author}{A.~M. Bruce}, \bibinfo{author}{E.~Casarejos},
  \bibinfo{author}{B.~Cederwall}, \bibinfo{author}{F.~C.~L. Crespi},
  \bibinfo{author}{B.~Hadinia}, \bibinfo{author}{M.~Hellstr\"om},
  \bibinfo{author}{R.~Hoischen}, \bibinfo{author}{G.~Ilie},
  \bibinfo{author}{J.~Jolie}, \bibinfo{author}{A.~Khaplanov},
  \bibinfo{author}{M.~Kmiecik}, \bibinfo{author}{R.~Kumar},
  \bibinfo{author}{A.~Maj}, \bibinfo{author}{S.~Mandal},
  \bibinfo{author}{F.~Montes}, \bibinfo{author}{S.~Myalski},
  \bibinfo{author}{G.~S. Simpson}, \bibinfo{author}{S.~J. Steer},
  \bibinfo{author}{S.~Tashenov}, \bibinfo{author}{O.~Wieland},
\newblock \bibinfo{title}{Observation of isomeric decays in the $r$-process
  waiting-point nucleus $^{130}\mathrm{Cd}_{82}$},
\newblock \bibinfo{journal}{Phys. Rev. Lett.} \bibinfo{volume}{99}
  (\bibinfo{year}{2007}) \bibinfo{pages}{132501}. \URLprefix
  \url{https://link.aps.org/doi/10.1103/PhysRevLett.99.132501}.
  \DOIprefix\doi{10.1103/PhysRevLett.99.132501}.
\bibitem[{Watanabe et~al.(2013)Watanabe, Lorusso, Nishimura, Xu, Sumikama,
  S\"oderstr\"om, Doornenbal, Browne, Gey, Jung, Taprogge, Vajta, Wu, Yagi,
  Baba, Benzoni, Chae, Crespi, Fukuda, Gernh\"auser, Inabe, Isobe, Jungclaus,
  Kameda, Kim, Kim, Kojouharov, Kondev, Kubo, Kurz, Kwon, Lane, Li, Moon,
  Montaner-Piz\'a, Moschner, Naqvi, Niikura, Nishibata, Nishimura, Odahara,
  Orlandi, Patel, Podoly\'ak, Sakurai, Schaffner, Simpson, Steiger, Suzuki,
  Takeda, Wendt, and Yoshinaga}]{PhysRevLett.111.152501}
\bibinfo{author}{H.~Watanabe}, \bibinfo{author}{G.~Lorusso},
  \bibinfo{author}{S.~Nishimura}, \bibinfo{author}{Z.~Y. Xu},
  \bibinfo{author}{T.~Sumikama}, \bibinfo{author}{P.-A. S\"oderstr\"om},
  \bibinfo{author}{P.~Doornenbal}, \bibinfo{author}{F.~Browne},
  \bibinfo{author}{G.~Gey}, \bibinfo{author}{H.~S. Jung},
  \bibinfo{author}{J.~Taprogge}, \bibinfo{author}{Z.~Vajta},
  \bibinfo{author}{J.~Wu}, \bibinfo{author}{A.~Yagi},
  \bibinfo{author}{H.~Baba}, \bibinfo{author}{G.~Benzoni},
  \bibinfo{author}{K.~Y. Chae}, \bibinfo{author}{F.~C.~L. Crespi},
  \bibinfo{author}{N.~Fukuda}, \bibinfo{author}{R.~Gernh\"auser},
  \bibinfo{author}{N.~Inabe}, \bibinfo{author}{T.~Isobe},
  \bibinfo{author}{A.~Jungclaus}, \bibinfo{author}{D.~Kameda},
  \bibinfo{author}{G.~D. Kim}, \bibinfo{author}{Y.~K. Kim},
  \bibinfo{author}{I.~Kojouharov}, \bibinfo{author}{F.~G. Kondev},
  \bibinfo{author}{T.~Kubo}, \bibinfo{author}{N.~Kurz}, \bibinfo{author}{Y.~K.
  Kwon}, \bibinfo{author}{G.~J. Lane}, \bibinfo{author}{Z.~Li},
  \bibinfo{author}{C.-B. Moon}, \bibinfo{author}{A.~Montaner-Piz\'a},
  \bibinfo{author}{K.~Moschner}, \bibinfo{author}{F.~Naqvi},
  \bibinfo{author}{M.~Niikura}, \bibinfo{author}{H.~Nishibata},
  \bibinfo{author}{D.~Nishimura}, \bibinfo{author}{A.~Odahara},
  \bibinfo{author}{R.~Orlandi}, \bibinfo{author}{Z.~Patel},
  \bibinfo{author}{Z.~Podoly\'ak}, \bibinfo{author}{H.~Sakurai},
  \bibinfo{author}{H.~Schaffner}, \bibinfo{author}{G.~S. Simpson},
  \bibinfo{author}{K.~Steiger}, \bibinfo{author}{H.~Suzuki},
  \bibinfo{author}{H.~Takeda}, \bibinfo{author}{A.~Wendt},
  \bibinfo{author}{K.~Yoshinaga},
\newblock \bibinfo{title}{Isomers in $^{128}\mathrm{Pd}$ and
  $^{126}\mathrm{Pd}$: Evidence for a robust shell closure at the neutron magic
  number 82 in exotic palladium isotopes},
\newblock \bibinfo{journal}{Phys. Rev. Lett.} \bibinfo{volume}{111}
  (\bibinfo{year}{2013}) \bibinfo{pages}{152501}. \URLprefix
  \url{https://link.aps.org/doi/10.1103/PhysRevLett.111.152501}.
  \DOIprefix\doi{10.1103/PhysRevLett.111.152501}.
\bibitem[{McNeill et~al.(1989)McNeill, Blomqvist, Chishti, Daly, Gelletly,
  Hotchkis, Piiparinen, Varley, and Woods}]{PhysRevLett.63.860}
\bibinfo{author}{J.~H. McNeill}, \bibinfo{author}{J.~Blomqvist},
  \bibinfo{author}{A.~A. Chishti}, \bibinfo{author}{P.~J. Daly},
  \bibinfo{author}{W.~Gelletly}, \bibinfo{author}{M.~A.~C. Hotchkis},
  \bibinfo{author}{M.~Piiparinen}, \bibinfo{author}{B.~J. Varley},
  \bibinfo{author}{P.~J. Woods},
\newblock \bibinfo{title}{Exotic n=82 nuclei $^{153}\mathrm{Lu}$ and
  $^{154}\mathrm{Hf}$ and filling of the \ensuremath{\pi}${h}_{11/2}$
  subshell},
\newblock \bibinfo{journal}{Phys. Rev. Lett.} \bibinfo{volume}{63}
  (\bibinfo{year}{1989}) \bibinfo{pages}{860--863}. \URLprefix
  \url{https://link.aps.org/doi/10.1103/PhysRevLett.63.860}.
  \DOIprefix\doi{10.1103/PhysRevLett.63.860}.
\bibitem[{Caurier et~al.(2003)Caurier, Rejmund, and Grawe}]{PhysRevC.67.054310}
\bibinfo{author}{E.~Caurier}, \bibinfo{author}{M.~Rejmund},
  \bibinfo{author}{H.~Grawe},
\newblock \bibinfo{title}{Large-scale shell model calculations for the $n=126$
  isotones {Po}--{Pu}},
\newblock \bibinfo{journal}{Phys. Rev. C} \bibinfo{volume}{67}
  (\bibinfo{year}{2003}) \bibinfo{pages}{054310}. \URLprefix
  \url{https://link.aps.org/doi/10.1103/PhysRevC.67.054310}.
  \DOIprefix\doi{10.1103/PhysRevC.67.054310}.
\bibitem[{Steer et~al.(2008)Steer, Podoly\'ak, Pietri, G\'orska, Regan,
  Rudolph, Werner-Malento, Garnsworthy, Hoischen, Gerl, Wollersheim, Maier,
  Grawe, Becker, Bednarczyk, C\'aceres, Doornenbal, Geissel, Gr\k{e}bosz,
  Kelic, Kojouharov, Kurz, Montes, Prokopowicz, Saito, Schaffner, Tashenov,
  Heinz, Pf\"utzner, Kurtukian-Nieto, Benzoni, Jungclaus, Balabanski, Brandau,
  Brown, Bruce, Catford, Cullen, Dombr\'adi, Estevez, Gelletly, Ilie, Jolie,
  Jones, Kmiecik, Kondev, Kr\"ucken, Lalkovski, Liu, Maj, Myalski, Schwertel,
  Shizuma, Walker, and Wieland}]{PhysRevC.78.061302}
\bibinfo{author}{S.~J. Steer}, \bibinfo{author}{Z.~Podoly\'ak},
  \bibinfo{author}{S.~Pietri}, \bibinfo{author}{M.~G\'orska},
  \bibinfo{author}{P.~H. Regan}, \bibinfo{author}{D.~Rudolph},
  \bibinfo{author}{E.~Werner-Malento}, \bibinfo{author}{A.~B. Garnsworthy},
  \bibinfo{author}{R.~Hoischen}, \bibinfo{author}{J.~Gerl},
  \bibinfo{author}{H.~J. Wollersheim}, \bibinfo{author}{K.~H. Maier},
  \bibinfo{author}{H.~Grawe}, \bibinfo{author}{F.~Becker},
  \bibinfo{author}{P.~Bednarczyk}, \bibinfo{author}{L.~C\'aceres},
  \bibinfo{author}{P.~Doornenbal}, \bibinfo{author}{H.~Geissel},
  \bibinfo{author}{J.~Gr\k{e}bosz}, \bibinfo{author}{A.~Kelic},
  \bibinfo{author}{I.~Kojouharov}, \bibinfo{author}{N.~Kurz},
  \bibinfo{author}{F.~Montes}, \bibinfo{author}{W.~Prokopowicz},
  \bibinfo{author}{T.~Saito}, \bibinfo{author}{H.~Schaffner},
  \bibinfo{author}{S.~Tashenov}, \bibinfo{author}{A.~Heinz},
  \bibinfo{author}{M.~Pf\"utzner}, \bibinfo{author}{T.~Kurtukian-Nieto},
  \bibinfo{author}{G.~Benzoni}, \bibinfo{author}{A.~Jungclaus},
  \bibinfo{author}{D.~L. Balabanski}, \bibinfo{author}{C.~Brandau},
  \bibinfo{author}{B.~A. Brown}, \bibinfo{author}{A.~M. Bruce},
  \bibinfo{author}{W.~N. Catford}, \bibinfo{author}{I.~J. Cullen},
  \bibinfo{author}{Z.~Dombr\'adi}, \bibinfo{author}{M.~E. Estevez},
  \bibinfo{author}{W.~Gelletly}, \bibinfo{author}{G.~Ilie},
  \bibinfo{author}{J.~Jolie}, \bibinfo{author}{G.~A. Jones},
  \bibinfo{author}{M.~Kmiecik}, \bibinfo{author}{F.~G. Kondev},
  \bibinfo{author}{R.~Kr\"ucken}, \bibinfo{author}{S.~Lalkovski},
  \bibinfo{author}{Z.~Liu}, \bibinfo{author}{A.~Maj},
  \bibinfo{author}{S.~Myalski}, \bibinfo{author}{S.~Schwertel},
  \bibinfo{author}{T.~Shizuma}, \bibinfo{author}{P.~M. Walker},
  \bibinfo{author}{O.~Wieland},
\newblock \bibinfo{title}{Single-particle behavior at ${N}=126$: Isomeric
  decays in neutron-rich $^{204}\mathrm{Pt}$},
\newblock \bibinfo{journal}{Phys. Rev. C} \bibinfo{volume}{78}
  (\bibinfo{year}{2008}) \bibinfo{pages}{061302}. \URLprefix
  \url{https://link.aps.org/doi/10.1103/PhysRevC.78.061302}.
  \DOIprefix\doi{10.1103/PhysRevC.78.061302}.
\bibitem[{Marchi et~al.(2014)Marchi, de~Angelis, Valiente-Dob\'on, Bader,
  Baugher, Bazin, Berryman, Bonaccorso, Clark, Coraggio, Crawford, Doncel,
  Farnea, Gade, Gadea, Gargano, Glasmacher, Gottardo, Gramegna, Itaco, John,
  Kumar, Lenzi, Lunardi, McDaniel, Michelagnoli, Mengoni, Modamio, Napoli,
  Quintana, Ratkiewicz, Recchia, Sahin, Stroberg, Weisshaar, Wimmer, and
  Winkler}]{PhysRevLett.113.182501}
\bibinfo{author}{T.~Marchi}, \bibinfo{author}{G.~de~Angelis},
  \bibinfo{author}{J.~J. Valiente-Dob\'on}, \bibinfo{author}{V.~M. Bader},
  \bibinfo{author}{T.~Baugher}, \bibinfo{author}{D.~Bazin},
  \bibinfo{author}{J.~Berryman}, \bibinfo{author}{A.~Bonaccorso},
  \bibinfo{author}{R.~Clark}, \bibinfo{author}{L.~Coraggio},
  \bibinfo{author}{H.~L. Crawford}, \bibinfo{author}{M.~Doncel},
  \bibinfo{author}{E.~Farnea}, \bibinfo{author}{A.~Gade},
  \bibinfo{author}{A.~Gadea}, \bibinfo{author}{A.~Gargano},
  \bibinfo{author}{T.~Glasmacher}, \bibinfo{author}{A.~Gottardo},
  \bibinfo{author}{F.~Gramegna}, \bibinfo{author}{N.~Itaco},
  \bibinfo{author}{P.~R. John}, \bibinfo{author}{R.~Kumar},
  \bibinfo{author}{S.~M. Lenzi}, \bibinfo{author}{S.~Lunardi},
  \bibinfo{author}{S.~McDaniel}, \bibinfo{author}{C.~Michelagnoli},
  \bibinfo{author}{D.~Mengoni}, \bibinfo{author}{V.~Modamio},
  \bibinfo{author}{D.~R. Napoli}, \bibinfo{author}{B.~Quintana},
  \bibinfo{author}{A.~Ratkiewicz}, \bibinfo{author}{F.~Recchia},
  \bibinfo{author}{E.~Sahin}, \bibinfo{author}{R.~Stroberg},
  \bibinfo{author}{D.~Weisshaar}, \bibinfo{author}{K.~Wimmer},
  \bibinfo{author}{R.~Winkler},
\newblock \bibinfo{title}{Quadrupole transition strength in the
  $^{74}\mathrm{Ni}$ nucleus and core polarization effects in the neutron-rich
  {Ni} isotopes},
\newblock \bibinfo{journal}{Phys. Rev. Lett.} \bibinfo{volume}{113}
  (\bibinfo{year}{2014}) \bibinfo{pages}{182501}. \URLprefix
  \url{https://link.aps.org/doi/10.1103/PhysRevLett.113.182501}.
  \DOIprefix\doi{10.1103/PhysRevLett.113.182501}.
\bibitem[{Kolos et~al.(2016)Kolos, Miller, Grzywacz, Iwasaki, Al-Shudifat,
  Bazin, Bingham, Braunroth, Cerizza, Gade, Lemasson, Liddick, Madurga, Morse,
  Portillo, Rajabali, Recchia, Riedinger, Voss, Walters, Weisshaar, Whitmore,
  Wimmer, and Tostevin}]{PhysRevLett.116.122502}
\bibinfo{author}{K.~Kolos}, \bibinfo{author}{D.~Miller},
  \bibinfo{author}{R.~Grzywacz}, \bibinfo{author}{H.~Iwasaki},
  \bibinfo{author}{M.~Al-Shudifat}, \bibinfo{author}{D.~Bazin},
  \bibinfo{author}{C.~R. Bingham}, \bibinfo{author}{T.~Braunroth},
  \bibinfo{author}{G.~Cerizza}, \bibinfo{author}{A.~Gade},
  \bibinfo{author}{A.~Lemasson}, \bibinfo{author}{S.~N. Liddick},
  \bibinfo{author}{M.~Madurga}, \bibinfo{author}{C.~Morse},
  \bibinfo{author}{M.~Portillo}, \bibinfo{author}{M.~M. Rajabali},
  \bibinfo{author}{F.~Recchia}, \bibinfo{author}{L.~L. Riedinger},
  \bibinfo{author}{P.~Voss}, \bibinfo{author}{W.~B. Walters},
  \bibinfo{author}{D.~Weisshaar}, \bibinfo{author}{K.~Whitmore},
  \bibinfo{author}{K.~Wimmer}, \bibinfo{author}{J.~A. Tostevin},
\newblock \bibinfo{title}{Direct lifetime measurements of the excited states in
  $^{72}\mathrm{Ni}$},
\newblock \bibinfo{journal}{Phys. Rev. Lett.} \bibinfo{volume}{116}
  (\bibinfo{year}{2016}) \bibinfo{pages}{122502}. \URLprefix
  \url{https://link.aps.org/doi/10.1103/PhysRevLett.116.122502}.
  \DOIprefix\doi{10.1103/PhysRevLett.116.122502}.
\bibitem[{Simpson et~al.(2014)Simpson, Gey, Jungclaus, Taprogge, Nishimura,
  Sieja, Doornenbal, Lorusso, S\"oderstr\"om, Sumikama, Xu, Baba, Browne,
  Fukuda, Inabe, Isobe, Jung, Kameda, Kim, Kim, Kojouharov, Kubo, Kurz, Kwon,
  Li, Sakurai, Schaffner, Shimizu, Suzuki, Takeda, Vajta, Watanabe, Wu, Yagi,
  Yoshinaga, B\"onig, Daugas, Drouet, Gernh\"auser, Ilieva, Kr\"oll,
  Montaner-Piz\'a, Moschner, M\"ucher, Na\"{\i}dja, Nishibata, Nowacki,
  Odahara, Orlandi, Steiger, and Wendt}]{PhysRevLett.113.132502}
\bibinfo{author}{G.~S. Simpson}, \bibinfo{author}{G.~Gey},
  \bibinfo{author}{A.~Jungclaus}, \bibinfo{author}{J.~Taprogge},
  \bibinfo{author}{S.~Nishimura}, \bibinfo{author}{K.~Sieja},
  \bibinfo{author}{P.~Doornenbal}, \bibinfo{author}{G.~Lorusso},
  \bibinfo{author}{P.-A. S\"oderstr\"om}, \bibinfo{author}{T.~Sumikama},
  \bibinfo{author}{Z.~Y. Xu}, \bibinfo{author}{H.~Baba},
  \bibinfo{author}{F.~Browne}, \bibinfo{author}{N.~Fukuda},
  \bibinfo{author}{N.~Inabe}, \bibinfo{author}{T.~Isobe},
  \bibinfo{author}{H.~S. Jung}, \bibinfo{author}{D.~Kameda},
  \bibinfo{author}{G.~D. Kim}, \bibinfo{author}{Y.-K. Kim},
  \bibinfo{author}{I.~Kojouharov}, \bibinfo{author}{T.~Kubo},
  \bibinfo{author}{N.~Kurz}, \bibinfo{author}{Y.~K. Kwon},
  \bibinfo{author}{Z.~Li}, \bibinfo{author}{H.~Sakurai},
  \bibinfo{author}{H.~Schaffner}, \bibinfo{author}{Y.~Shimizu},
  \bibinfo{author}{H.~Suzuki}, \bibinfo{author}{H.~Takeda},
  \bibinfo{author}{Z.~Vajta}, \bibinfo{author}{H.~Watanabe},
  \bibinfo{author}{J.~Wu}, \bibinfo{author}{A.~Yagi},
  \bibinfo{author}{K.~Yoshinaga}, \bibinfo{author}{S.~B\"onig},
  \bibinfo{author}{J.-M. Daugas}, \bibinfo{author}{F.~Drouet},
  \bibinfo{author}{R.~Gernh\"auser}, \bibinfo{author}{S.~Ilieva},
  \bibinfo{author}{T.~Kr\"oll}, \bibinfo{author}{A.~Montaner-Piz\'a},
  \bibinfo{author}{K.~Moschner}, \bibinfo{author}{D.~M\"ucher},
  \bibinfo{author}{H.~Na\"{\i}dja}, \bibinfo{author}{H.~Nishibata},
  \bibinfo{author}{F.~Nowacki}, \bibinfo{author}{A.~Odahara},
  \bibinfo{author}{R.~Orlandi}, \bibinfo{author}{K.~Steiger},
  \bibinfo{author}{A.~Wendt},
\newblock \bibinfo{title}{Yrast 6${}^{+}$ seniority isomers of
  ${}^{136,138}${Sn}},
\newblock \bibinfo{journal}{Phys. Rev. Lett.} \bibinfo{volume}{113}
  (\bibinfo{year}{2014}) \bibinfo{pages}{132502}. \URLprefix
  \url{https://link.aps.org/doi/10.1103/PhysRevLett.113.132502}.
  \DOIprefix\doi{10.1103/PhysRevLett.113.132502}.
\bibitem[{Gottardo et~al.(2012)Gottardo, Valiente-Dob\'on, Benzoni, Nicolini,
  Gadea, Lunardi, Boutachkov, Bruce, G\'orska, Grebosz, Pietri, Podoly\'ak,
  Pf\"utzner, Regan, Weick, Alc\'antara N\'u\~nez, Algora, Al-Dahan,
  de~Angelis, Ayyad, Alkhomashi, Allegro, Bazzacco, Benlliure, Bowry, Bracco,
  Bunce, Camera, Casarejos, Cortes, Crespi, Corsi, Denis~Bacelar, Deo,
  Domingo-Pardo, Doncel, Dombradi, Engert, Eppinger, Farrelly, Farinon, Farnea,
  Geissel, Gerl, Goel, Gregor, Habermann, Hoischen, Janik, Klupp, Kojouharov,
  Kurz, Lenzi, Leoni, Mandal, Menegazzo, Mengoni, Million, Morales, Napoli,
  Naqvi, Nociforo, Prochazka, Prokopowicz, Recchia, Ribas, Reed, Rudolph,
  Sahin, Schaffner, Sharma, Sitar, Siwal, Steiger, Strmen, Swan, Szarka, Ur,
  Walker, Wieland, Wollersheim, Nowacki, Maglione, and
  Zuker}]{PhysRevLett.109.162502}
\bibinfo{author}{A.~Gottardo}, \bibinfo{author}{J.~J. Valiente-Dob\'on},
  \bibinfo{author}{G.~Benzoni}, \bibinfo{author}{R.~Nicolini},
  \bibinfo{author}{A.~Gadea}, \bibinfo{author}{S.~Lunardi},
  \bibinfo{author}{P.~Boutachkov}, \bibinfo{author}{A.~M. Bruce},
  \bibinfo{author}{M.~G\'orska}, \bibinfo{author}{J.~Grebosz},
  \bibinfo{author}{S.~Pietri}, \bibinfo{author}{Z.~Podoly\'ak},
  \bibinfo{author}{M.~Pf\"utzner}, \bibinfo{author}{P.~H. Regan},
  \bibinfo{author}{H.~Weick}, \bibinfo{author}{J.~Alc\'antara N\'u\~nez},
  \bibinfo{author}{A.~Algora}, \bibinfo{author}{N.~Al-Dahan},
  \bibinfo{author}{G.~de~Angelis}, \bibinfo{author}{Y.~Ayyad},
  \bibinfo{author}{N.~Alkhomashi}, \bibinfo{author}{P.~R.~P. Allegro},
  \bibinfo{author}{D.~Bazzacco}, \bibinfo{author}{J.~Benlliure},
  \bibinfo{author}{M.~Bowry}, \bibinfo{author}{A.~Bracco},
  \bibinfo{author}{M.~Bunce}, \bibinfo{author}{F.~Camera},
  \bibinfo{author}{E.~Casarejos}, \bibinfo{author}{M.~L. Cortes},
  \bibinfo{author}{F.~C.~L. Crespi}, \bibinfo{author}{A.~Corsi},
  \bibinfo{author}{A.~M. Denis~Bacelar}, \bibinfo{author}{A.~Y. Deo},
  \bibinfo{author}{C.~Domingo-Pardo}, \bibinfo{author}{M.~Doncel},
  \bibinfo{author}{Z.~Dombradi}, \bibinfo{author}{T.~Engert},
  \bibinfo{author}{K.~Eppinger}, \bibinfo{author}{G.~F. Farrelly},
  \bibinfo{author}{F.~Farinon}, \bibinfo{author}{E.~Farnea},
  \bibinfo{author}{H.~Geissel}, \bibinfo{author}{J.~Gerl},
  \bibinfo{author}{N.~Goel}, \bibinfo{author}{E.~Gregor},
  \bibinfo{author}{T.~Habermann}, \bibinfo{author}{R.~Hoischen},
  \bibinfo{author}{R.~Janik}, \bibinfo{author}{S.~Klupp},
  \bibinfo{author}{I.~Kojouharov}, \bibinfo{author}{N.~Kurz},
  \bibinfo{author}{S.~M. Lenzi}, \bibinfo{author}{S.~Leoni},
  \bibinfo{author}{S.~Mandal}, \bibinfo{author}{R.~Menegazzo},
  \bibinfo{author}{D.~Mengoni}, \bibinfo{author}{B.~Million},
  \bibinfo{author}{A.~I. Morales}, \bibinfo{author}{D.~R. Napoli},
  \bibinfo{author}{F.~Naqvi}, \bibinfo{author}{C.~Nociforo},
  \bibinfo{author}{A.~Prochazka}, \bibinfo{author}{W.~Prokopowicz},
  \bibinfo{author}{F.~Recchia}, \bibinfo{author}{R.~V. Ribas},
  \bibinfo{author}{M.~W. Reed}, \bibinfo{author}{D.~Rudolph},
  \bibinfo{author}{E.~Sahin}, \bibinfo{author}{H.~Schaffner},
  \bibinfo{author}{A.~Sharma}, \bibinfo{author}{B.~Sitar},
  \bibinfo{author}{D.~Siwal}, \bibinfo{author}{K.~Steiger},
  \bibinfo{author}{P.~Strmen}, \bibinfo{author}{T.~P.~D. Swan},
  \bibinfo{author}{I.~Szarka}, \bibinfo{author}{C.~A. Ur},
  \bibinfo{author}{P.~M. Walker}, \bibinfo{author}{O.~Wieland},
  \bibinfo{author}{H.-J. Wollersheim}, \bibinfo{author}{F.~Nowacki},
  \bibinfo{author}{E.~Maglione}, \bibinfo{author}{A.~P. Zuker},
\newblock \bibinfo{title}{New isomers in the full seniority scheme of
  neutron-rich lead isotopes: The role of effective three-body forces},
\newblock \bibinfo{journal}{Phys. Rev. Lett.} \bibinfo{volume}{109}
  (\bibinfo{year}{2012}) \bibinfo{pages}{162502}. \URLprefix
  \url{https://link.aps.org/doi/10.1103/PhysRevLett.109.162502}.
  \DOIprefix\doi{10.1103/PhysRevLett.109.162502}.
\bibitem[{Walker and
  Podoly{\'a}k(2020{\natexlab{a}})}]{Walker_Phys.Scr.95.044004(2020)}
\bibinfo{author}{P.~M. Walker}, \bibinfo{author}{Z.~Podoly{\'a}k},
\newblock \bibinfo{title}{100 years of nuclear isomers—then and now},
\newblock \bibinfo{journal}{Phys. Scr.} \bibinfo{volume}{95}
  (\bibinfo{year}{2020}{\natexlab{a}}) \bibinfo{pages}{044004}. \URLprefix
  \url{https://doi.org/10.1088/1402-4896/ab635d}.
  \DOIprefix\doi{10.1088/1402-4896/ab635d}.
\bibitem[{Walker and Podoly{\'a}k(2020{\natexlab{b}})}]{Walker_Book_2020}
\bibinfo{author}{P.~M. Walker}, \bibinfo{author}{Z.~Podoly{\'a}k},
  \bibinfo{title}{Nuclear Isomers}, \bibinfo{publisher}{Springer Nature
  Singapore}, \bibinfo{address}{Singapore}, \bibinfo{year}{2020}{\natexlab{b}},
  pp. \bibinfo{pages}{1--37}. \URLprefix
  \url{https://doi.org/10.1007/978-981-15-8818-1_46-1}.
  \DOIprefix\doi{10.1007/978-981-15-8818-1_46-1}.
\bibitem[{Mach et~al.(2017)Mach, Korgul, G\'orska, Grawe, Matea,
  St\ifmmode~\u{a}\else \u{a}\fi{}noiu, Fraile, Penionzkevich, Santos, Verney,
  Lukyanov, Cederwall, Covello, Dlouh\'y, Fogelberg, De~France, Gargano,
  Georgiev, Grzywacz, Lisetskiy, Mrazek, Nowacki, P\l{}\'ociennik, Podoly\'ak,
  Ray, Ruchowska, Saint-Laurent, Sawicka, Stodel, and
  Tarasov}]{PhysRevC.95.014313}
\bibinfo{author}{H.~Mach}, \bibinfo{author}{A.~Korgul},
  \bibinfo{author}{M.~G\'orska}, \bibinfo{author}{H.~Grawe},
  \bibinfo{author}{I.~Matea}, \bibinfo{author}{M.~St\ifmmode~\u{a}\else
  \u{a}\fi{}noiu}, \bibinfo{author}{L.~M. Fraile}, \bibinfo{author}{Y.~E.
  Penionzkevich}, \bibinfo{author}{F.~D.~O. Santos},
  \bibinfo{author}{D.~Verney}, \bibinfo{author}{S.~Lukyanov},
  \bibinfo{author}{B.~Cederwall}, \bibinfo{author}{A.~Covello},
  \bibinfo{author}{Z.~Dlouh\'y}, \bibinfo{author}{B.~Fogelberg},
  \bibinfo{author}{G.~De~France}, \bibinfo{author}{A.~Gargano},
  \bibinfo{author}{G.~Georgiev}, \bibinfo{author}{R.~Grzywacz},
  \bibinfo{author}{A.~F. Lisetskiy}, \bibinfo{author}{J.~Mrazek},
  \bibinfo{author}{F.~Nowacki}, \bibinfo{author}{W.~A. P\l{}\'ociennik},
  \bibinfo{author}{Z.~Podoly\'ak}, \bibinfo{author}{S.~Ray},
  \bibinfo{author}{E.~Ruchowska}, \bibinfo{author}{M.-G. Saint-Laurent},
  \bibinfo{author}{M.~Sawicka}, \bibinfo{author}{C.~Stodel},
  \bibinfo{author}{O.~Tarasov},
\newblock \bibinfo{title}{Ultrafast-timing lifetime measurements in
  $^{94}\mathrm{Ru}$ and $^{96}\mathrm{Pd}$: Breakdown of the seniority scheme
  in ${N}=50$ isotones},
\newblock \bibinfo{journal}{Phys. Rev. C} \bibinfo{volume}{95}
  (\bibinfo{year}{2017}) \bibinfo{pages}{014313}. \URLprefix
  \url{https://link.aps.org/doi/10.1103/PhysRevC.95.014313}.
  \DOIprefix\doi{10.1103/PhysRevC.95.014313}.
\bibitem[{Das et~al.(2022)Das, Cederwall, Qi, G\'orska, Regan, Aktas, Albers,
  Banerjee, Chishti, Gerl, Hubbard, Jazrawi, Jolie, Mistry, Polettini, Yaneva,
  Alhomaidhi, Zhao, Arici, Bagchi, Benzoni, Boutachkov, Davinson, Dickel,
  Haettner, Hall, Hornung, Hucka, John, Kojouharov, Kn\"obel, Kostyleva,
  Kuzminchuk, Mukha, Plass, Nara~Singh, Vasiljevi\ifmmode~\acute{c}\else
  \'{c}\fi{}, Pietri, Podoly\'ak, Rudigier, R\"osch, Sahin, Schaffner,
  Scheidenberger, Schirru, Sharma, Shearman, Tanaka,
  Vesi\ifmmode~\acute{c}\else \'{c}\fi{}, Weick, Wollersheim, Ahmed, Algora,
  Appleton, Benito, Blazhev, Bracco, Bruce, Brunet, Canavan, Esmaylzadeh,
  Fraile, H\"afner, Heggen, Kahl, Karayonchev, Kern, Korgul, Kosir, Kurz,
  Lozeva, Mikolajczuk, Napiralla, Page, Petrache, Pietralla, R\'egis,
  Ruotsalainen, Sexton, Sanchez-Temble, Si, Vilhena, Werner, Wiederhold, Witt,
  Woods, and Zimba}]{PhysRevC.105.L031304}
\bibinfo{author}{B.~Das}, \bibinfo{author}{B.~Cederwall},
  \bibinfo{author}{C.~Qi}, \bibinfo{author}{M.~G\'orska},
  \bibinfo{author}{P.~H. Regan}, \bibinfo{author}{O.~Aktas},
  \bibinfo{author}{H.~M. Albers}, \bibinfo{author}{A.~Banerjee},
  \bibinfo{author}{M.~M.~R. Chishti}, \bibinfo{author}{J.~Gerl},
  \bibinfo{author}{N.~Hubbard}, \bibinfo{author}{S.~Jazrawi},
  \bibinfo{author}{J.~Jolie}, \bibinfo{author}{A.~K. Mistry},
  \bibinfo{author}{M.~Polettini}, \bibinfo{author}{A.~Yaneva},
  \bibinfo{author}{S.~Alhomaidhi}, \bibinfo{author}{J.~Zhao},
  \bibinfo{author}{T.~Arici}, \bibinfo{author}{S.~Bagchi},
  \bibinfo{author}{G.~Benzoni}, \bibinfo{author}{P.~Boutachkov},
  \bibinfo{author}{T.~Davinson}, \bibinfo{author}{T.~Dickel},
  \bibinfo{author}{E.~Haettner}, \bibinfo{author}{O.~Hall},
  \bibinfo{author}{C.~Hornung}, \bibinfo{author}{J.~P. Hucka},
  \bibinfo{author}{P.~R. John}, \bibinfo{author}{I.~Kojouharov},
  \bibinfo{author}{R.~Kn\"obel}, \bibinfo{author}{D.~Kostyleva},
  \bibinfo{author}{N.~Kuzminchuk}, \bibinfo{author}{I.~Mukha},
  \bibinfo{author}{W.~R. Plass}, \bibinfo{author}{B.~S. Nara~Singh},
  \bibinfo{author}{J.~Vasiljevi\ifmmode~\acute{c}\else \'{c}\fi{}},
  \bibinfo{author}{S.~Pietri}, \bibinfo{author}{Z.~Podoly\'ak},
  \bibinfo{author}{M.~Rudigier}, \bibinfo{author}{H.~R\"osch},
  \bibinfo{author}{E.~Sahin}, \bibinfo{author}{H.~Schaffner},
  \bibinfo{author}{C.~Scheidenberger}, \bibinfo{author}{F.~Schirru},
  \bibinfo{author}{A.~Sharma}, \bibinfo{author}{R.~Shearman},
  \bibinfo{author}{Y.~Tanaka}, \bibinfo{author}{J.~Vesi\ifmmode~\acute{c}\else
  \'{c}\fi{}}, \bibinfo{author}{H.~Weick}, \bibinfo{author}{H.~J. Wollersheim},
  \bibinfo{author}{U.~Ahmed}, \bibinfo{author}{A.~Algora},
  \bibinfo{author}{C.~Appleton}, \bibinfo{author}{J.~Benito},
  \bibinfo{author}{A.~Blazhev}, \bibinfo{author}{A.~Bracco},
  \bibinfo{author}{A.~M. Bruce}, \bibinfo{author}{M.~Brunet},
  \bibinfo{author}{R.~Canavan}, \bibinfo{author}{A.~Esmaylzadeh},
  \bibinfo{author}{L.~M. Fraile}, \bibinfo{author}{G.~H\"afner},
  \bibinfo{author}{H.~Heggen}, \bibinfo{author}{D.~Kahl},
  \bibinfo{author}{V.~Karayonchev}, \bibinfo{author}{R.~Kern},
  \bibinfo{author}{A.~Korgul}, \bibinfo{author}{G.~Kosir},
  \bibinfo{author}{N.~Kurz}, \bibinfo{author}{R.~Lozeva},
  \bibinfo{author}{M.~Mikolajczuk}, \bibinfo{author}{P.~Napiralla},
  \bibinfo{author}{R.~Page}, \bibinfo{author}{C.~M. Petrache},
  \bibinfo{author}{N.~Pietralla}, \bibinfo{author}{J.-M. R\'egis},
  \bibinfo{author}{P.~Ruotsalainen}, \bibinfo{author}{L.~Sexton},
  \bibinfo{author}{V.~Sanchez-Temble}, \bibinfo{author}{M.~Si},
  \bibinfo{author}{J.~Vilhena}, \bibinfo{author}{V.~Werner},
  \bibinfo{author}{J.~Wiederhold}, \bibinfo{author}{W.~Witt},
  \bibinfo{author}{P.~J. Woods}, \bibinfo{author}{G.~Zimba},
\newblock \bibinfo{title}{Nature of seniority symmetry breaking in the
  semimagic nucleus $^{94}\mathrm{Ru}$},
\newblock \bibinfo{journal}{Phys. Rev. C} \bibinfo{volume}{105}
  (\bibinfo{year}{2022}) \bibinfo{pages}{L031304}. \URLprefix
  \url{https://link.aps.org/doi/10.1103/PhysRevC.105.L031304}.
  \DOIprefix\doi{10.1103/PhysRevC.105.L031304}.
\bibitem[{Das et~al.(2024)Das, Cederwall, Qi, G\'orska, Regan, Aktas, Albers,
  Banerjee, Chishti, Gerl, Hubbard, Jazrawi, Jolie, Mistry, Nowacki, Polettini,
  Yaneva, Ahmed, Alhomaidhi, Algora, Appleton, Arici, Bagchi, Benzoni, Benito,
  Blazhev, Boutachkov, Bracco, Bruce, Brunet, Canavan, Davinson, Dickel,
  Esmaylzadeh, Fraile, Haettner, Hall, H\"afner, Heggen, Hornung, Hucka, John,
  Kahl, Karayonchev, Kern, Kn\"obel, Korgul, Kosir, Kojouharov, Kostyleva,
  Kuzminchuk, Kurz, Liotta, Lozeva, Mikolajczuk, Mukha, Napiralla, Page,
  Petrache, Pietralla, Pietri, Pla\ss{}, Podoly\'ak, R\'egis, Rudigier,
  R\"osch, Ruotsalainen, Sahin, S\'anchez-Tembleque, Schaffner, Scheidenberger,
  Schirru, Sexton, Singh, Sharma, Shearman, Si, Tanaka,
  Vasiljevi\ifmmode~\acute{c}\else \'{c}\fi{}, Vesi\ifmmode~\acute{c}\else
  \'{c}\fi{}, Vilhena, Weick, Wollersheim, Werner, Wiederhold, Witt, Woods,
  Zimba, and Zhao}]{PhysRevResearch.6.L022038}
\bibinfo{author}{B.~Das}, \bibinfo{author}{B.~Cederwall},
  \bibinfo{author}{C.~Qi}, \bibinfo{author}{M.~G\'orska},
  \bibinfo{author}{P.~H. Regan}, \bibinfo{author}{O.~Aktas},
  \bibinfo{author}{H.~M. Albers}, \bibinfo{author}{A.~Banerjee},
  \bibinfo{author}{M.~M.~R. Chishti}, \bibinfo{author}{J.~Gerl},
  \bibinfo{author}{N.~Hubbard}, \bibinfo{author}{S.~Jazrawi},
  \bibinfo{author}{J.~Jolie}, \bibinfo{author}{A.~K. Mistry},
  \bibinfo{author}{F.~Nowacki}, \bibinfo{author}{M.~Polettini},
  \bibinfo{author}{A.~Yaneva}, \bibinfo{author}{U.~Ahmed},
  \bibinfo{author}{S.~Alhomaidhi}, \bibinfo{author}{A.~Algora},
  \bibinfo{author}{C.~Appleton}, \bibinfo{author}{T.~Arici},
  \bibinfo{author}{S.~Bagchi}, \bibinfo{author}{G.~Benzoni},
  \bibinfo{author}{J.~Benito}, \bibinfo{author}{A.~Blazhev},
  \bibinfo{author}{P.~Boutachkov}, \bibinfo{author}{A.~Bracco},
  \bibinfo{author}{A.~M. Bruce}, \bibinfo{author}{M.~Brunet},
  \bibinfo{author}{R.~Canavan}, \bibinfo{author}{T.~Davinson},
  \bibinfo{author}{T.~Dickel}, \bibinfo{author}{A.~Esmaylzadeh},
  \bibinfo{author}{L.~M. Fraile}, \bibinfo{author}{E.~Haettner},
  \bibinfo{author}{O.~Hall}, \bibinfo{author}{G.~H\"afner},
  \bibinfo{author}{H.~Heggen}, \bibinfo{author}{C.~Hornung},
  \bibinfo{author}{J.~P. Hucka}, \bibinfo{author}{P.~R. John},
  \bibinfo{author}{D.~Kahl}, \bibinfo{author}{V.~Karayonchev},
  \bibinfo{author}{R.~Kern}, \bibinfo{author}{R.~Kn\"obel},
  \bibinfo{author}{A.~Korgul}, \bibinfo{author}{G.~Kosir},
  \bibinfo{author}{I.~Kojouharov}, \bibinfo{author}{D.~Kostyleva},
  \bibinfo{author}{N.~Kuzminchuk}, \bibinfo{author}{N.~Kurz},
  \bibinfo{author}{R.~Liotta}, \bibinfo{author}{R.~Lozeva},
  \bibinfo{author}{M.~Mikolajczuk}, \bibinfo{author}{I.~Mukha},
  \bibinfo{author}{P.~Napiralla}, \bibinfo{author}{R.~Page},
  \bibinfo{author}{C.~M. Petrache}, \bibinfo{author}{N.~Pietralla},
  \bibinfo{author}{S.~Pietri}, \bibinfo{author}{W.~R. Pla\ss{}},
  \bibinfo{author}{Z.~Podoly\'ak}, \bibinfo{author}{J.-M. R\'egis},
  \bibinfo{author}{M.~Rudigier}, \bibinfo{author}{H.~R\"osch},
  \bibinfo{author}{P.~Ruotsalainen}, \bibinfo{author}{E.~Sahin},
  \bibinfo{author}{V.~S\'anchez-Tembleque}, \bibinfo{author}{H.~Schaffner},
  \bibinfo{author}{C.~Scheidenberger}, \bibinfo{author}{F.~Schirru},
  \bibinfo{author}{L.~Sexton}, \bibinfo{author}{B.~S.~N. Singh},
  \bibinfo{author}{A.~Sharma}, \bibinfo{author}{R.~Shearman},
  \bibinfo{author}{M.~Si}, \bibinfo{author}{Y.~K. Tanaka},
  \bibinfo{author}{J.~Vasiljevi\ifmmode~\acute{c}\else \'{c}\fi{}},
  \bibinfo{author}{J.~Vesi\ifmmode~\acute{c}\else \'{c}\fi{}},
  \bibinfo{author}{J.~Vilhena}, \bibinfo{author}{H.~Weick},
  \bibinfo{author}{H.~J. Wollersheim}, \bibinfo{author}{V.~Werner},
  \bibinfo{author}{J.~Wiederhold}, \bibinfo{author}{W.~Witt},
  \bibinfo{author}{P.~J. Woods}, \bibinfo{author}{G.~Zimba},
  \bibinfo{author}{J.~Zhao},
\newblock \bibinfo{title}{Broken seniority symmetry in the semimagic proton
  mid-shell nucleus $^{95}\mathrm{Rh}$},
\newblock \bibinfo{journal}{Phys. Rev. Res.} \bibinfo{volume}{6}
  (\bibinfo{year}{2024}) \bibinfo{pages}{L022038}. \URLprefix
  \url{https://link.aps.org/doi/10.1103/PhysRevResearch.6.L022038}.
  \DOIprefix\doi{10.1103/PhysRevResearch.6.L022038}.
\bibitem[{Rowe and Rosensteel(2001)}]{PhysRevLett.87.172501}
\bibinfo{author}{D.~J. Rowe}, \bibinfo{author}{G.~Rosensteel},
\newblock \bibinfo{title}{Partially solvable pair-coupling models with
  seniority-conserving interactions},
\newblock \bibinfo{journal}{Phys. Rev. Lett.} \bibinfo{volume}{87}
  (\bibinfo{year}{2001}) \bibinfo{pages}{172501}. \URLprefix
  \url{https://link.aps.org/doi/10.1103/PhysRevLett.87.172501}.
  \DOIprefix\doi{10.1103/PhysRevLett.87.172501}.
\bibitem[{Lisetskiy et~al.(2004)Lisetskiy, Brown, Horoi, and
  Grawe}]{PhysRevC.70.044314}
\bibinfo{author}{A.~F. Lisetskiy}, \bibinfo{author}{B.~A. Brown},
  \bibinfo{author}{M.~Horoi}, \bibinfo{author}{H.~Grawe},
\newblock \bibinfo{title}{New ${T}=1$ effective interactions for the
  ${f}_{5/2}\phantom{\rule{0.3em}{0ex}}{p}_{3/2}\phantom{\rule{0.3em}{0ex}}{p}_{1/2}\phantom{\rule{0.3em}{0ex}}{g}_{9/2}$
  model space: Implications for valence-mirror symmetry and seniority isomers},
\newblock \bibinfo{journal}{Phys. Rev. C} \bibinfo{volume}{70}
  (\bibinfo{year}{2004}) \bibinfo{pages}{044314}. \URLprefix
  \url{https://link.aps.org/doi/10.1103/PhysRevC.70.044314}.
  \DOIprefix\doi{10.1103/PhysRevC.70.044314}.
\bibitem[{Qi(2011)}]{PhysRevC.83.014307}
\bibinfo{author}{C.~Qi},
\newblock \bibinfo{title}{Partial conservation of seniority in the $j=9/2$
  shell: Analytic and numerical studies},
\newblock \bibinfo{journal}{Phys. Rev. C} \bibinfo{volume}{83}
  (\bibinfo{year}{2011}) \bibinfo{pages}{014307}. \URLprefix
  \url{https://link.aps.org/doi/10.1103/PhysRevC.83.014307}.
  \DOIprefix\doi{10.1103/PhysRevC.83.014307}.
\bibitem[{Qian and Qi(2018)}]{PhysRevC.98.061303}
\bibinfo{author}{Y.~Qian}, \bibinfo{author}{C.~Qi},
\newblock \bibinfo{title}{Partial seniority conservation and solvability of
  single-$j$ systems},
\newblock \bibinfo{journal}{Phys. Rev. C} \bibinfo{volume}{98}
  (\bibinfo{year}{2018}) \bibinfo{pages}{061303}. \URLprefix
  \url{https://link.aps.org/doi/10.1103/PhysRevC.98.061303}.
  \DOIprefix\doi{10.1103/PhysRevC.98.061303}.
\bibitem[{Van~Isacker and Heinze(2008)}]{PhysRevLett.100.052501}
\bibinfo{author}{P.~Van~Isacker}, \bibinfo{author}{S.~Heinze},
\newblock \bibinfo{title}{Partial conservation of seniority and nuclear
  isomerism},
\newblock \bibinfo{journal}{Phys. Rev. Lett.} \bibinfo{volume}{100}
  (\bibinfo{year}{2008}) \bibinfo{pages}{052501}. \URLprefix
  \url{https://link.aps.org/doi/10.1103/PhysRevLett.100.052501}.
  \DOIprefix\doi{10.1103/PhysRevLett.100.052501}.
\bibitem[{Epelbaum et~al.(2009)Epelbaum, Hammer, and
  Mei\ss{}ner}]{E.Epelbaum_RevModPhys.81.1773(2009)}
\bibinfo{author}{E.~Epelbaum}, \bibinfo{author}{H.-W. Hammer},
  \bibinfo{author}{U.-G. Mei\ss{}ner},
\newblock \bibinfo{title}{Modern theory of nuclear forces},
\newblock \bibinfo{journal}{Rev. Mod. Phys.} \bibinfo{volume}{81}
  (\bibinfo{year}{2009}) \bibinfo{pages}{1773--1825}. \URLprefix
  \url{https://link.aps.org/doi/10.1103/RevModPhys.81.1773}.
  \DOIprefix\doi{10.1103/RevModPhys.81.1773}.
\bibitem[{Machleidt and Entem(2011)}]{R.Machleidt_Phys.Rep.503.1(2011)}
\bibinfo{author}{R.~Machleidt}, \bibinfo{author}{D.~Entem},
\newblock \bibinfo{title}{Chiral effective field theory and nuclear forces},
\newblock \bibinfo{journal}{Phys. Rep.} \bibinfo{volume}{503}
  (\bibinfo{year}{2011}) \bibinfo{pages}{1--75}. \URLprefix
  \url{https://www.sciencedirect.com/science/article/pii/S0370157311000457}.
  \DOIprefix\doi{https://doi.org/10.1016/j.physrep.2011.02.001}.
\bibitem[{Hergert(2020)}]{hergert2020}
\bibinfo{author}{H.~Hergert},
\newblock \bibinfo{title}{A guided tour of ab initio nuclear many-body theory},
\newblock \bibinfo{journal}{Front. Phys.} \bibinfo{volume}{8}
  (\bibinfo{year}{2020}) \bibinfo{pages}{379}. \URLprefix
  \url{https://www.frontiersin.org/article/10.3389/fphy.2020.00379}.
  \DOIprefix\doi{10.3389/fphy.2020.00379}.
\bibitem[{Barrett et~al.(2013)Barrett, Navrátil, and Vary}]{BARRETT2013131}
\bibinfo{author}{B.~R. Barrett}, \bibinfo{author}{P.~Navrátil},
  \bibinfo{author}{J.~P. Vary},
\newblock \bibinfo{title}{Ab initio no core shell model},
\newblock \bibinfo{journal}{Prog. Part. Nucl. Phys} \bibinfo{volume}{69}
  (\bibinfo{year}{2013}) \bibinfo{pages}{131--181}. \URLprefix
  \url{https://www.sciencedirect.com/science/article/pii/S0146641012001184}.
  \DOIprefix\doi{https://doi.org/10.1016/j.ppnp.2012.10.003}.
\bibitem[{Carlson et~al.(2015)Carlson, Gandolfi, Pederiva, Pieper, Schiavilla,
  Schmidt, and Wiringa}]{RevModPhys.87.1067}
\bibinfo{author}{J.~Carlson}, \bibinfo{author}{S.~Gandolfi},
  \bibinfo{author}{F.~Pederiva}, \bibinfo{author}{S.~C. Pieper},
  \bibinfo{author}{R.~Schiavilla}, \bibinfo{author}{K.~E. Schmidt},
  \bibinfo{author}{R.~B. Wiringa},
\newblock \bibinfo{title}{Quantum monte carlo methods for nuclear physics},
\newblock \bibinfo{journal}{Rev. Mod. Phys.} \bibinfo{volume}{87}
  (\bibinfo{year}{2015}) \bibinfo{pages}{1067--1118}. \URLprefix
  \url{https://link.aps.org/doi/10.1103/RevModPhys.87.1067}.
  \DOIprefix\doi{10.1103/RevModPhys.87.1067}.
\bibitem[{Hagen et~al.(2014)Hagen, Papenbrock, Hjorth-Jensen, and
  Dean}]{Hagen_2014}
\bibinfo{author}{G.~Hagen}, \bibinfo{author}{T.~Papenbrock},
  \bibinfo{author}{M.~Hjorth-Jensen}, \bibinfo{author}{D.~J. Dean},
\newblock \bibinfo{title}{Coupled-cluster computations of atomic nuclei},
\newblock \bibinfo{journal}{Rep. Prog. Phys} \bibinfo{volume}{77}
  (\bibinfo{year}{2014}) \bibinfo{pages}{096302}. \URLprefix
  \url{https://doi.org/10.1088/0034-4885/77/9/096302}.
  \DOIprefix\doi{10.1088/0034-4885/77/9/096302}.
\bibitem[{Hu et~al.(2019)Hu, Wu, Sun, and Xu}]{GIMSRG_PhysRevC.99.061302(2019)}
\bibinfo{author}{B.~S. Hu}, \bibinfo{author}{Q.~Wu}, \bibinfo{author}{Z.~H.
  Sun}, \bibinfo{author}{F.~R. Xu},
\newblock \bibinfo{title}{Ab initio gamow in-medium similarity renormalization
  group with resonance and continuum},
\newblock \bibinfo{journal}{Phys. Rev. C} \bibinfo{volume}{99}
  (\bibinfo{year}{2019}) \bibinfo{pages}{061302(R)}. \URLprefix
  \url{https://link.aps.org/doi/10.1103/PhysRevC.99.061302}.
  \DOIprefix\doi{10.1103/PhysRevC.99.061302}.
\bibitem[{Stroberg et~al.(2019)Stroberg, Hergert, Bogner, and
  Holt}]{doi:10.1146/annurev-nucl-101917-021120}
\bibinfo{author}{S.~R. Stroberg}, \bibinfo{author}{H.~Hergert},
  \bibinfo{author}{S.~K. Bogner}, \bibinfo{author}{J.~D. Holt},
\newblock \bibinfo{title}{Nonempirical interactions for the nuclear shell
  model: An update},
\newblock \bibinfo{journal}{Ann. Rev. Nucl. Part. Sci} \bibinfo{volume}{69}
  (\bibinfo{year}{2019}) \bibinfo{pages}{307--362}. \URLprefix
  \url{https://doi.org/10.1146/annurev-nucl-101917-021120}.
  \DOIprefix\doi{10.1146/annurev-nucl-101917-021120}.
  \href{http://arxiv.org/abs/https://doi.org/10.1146/annurev-nucl-101917-021120}{{\tt
  arXiv:https://doi.org/10.1146/annurev-nucl-101917-021120}}.
\bibitem[{Yuan et~al.(2022)Yuan, Fan, Hu, Li, Zhang, Wang, Sun, Ma, and
  Xu}]{DIMSRG_PhysRevC.105.L061303(2022)}
\bibinfo{author}{Q.~Yuan}, \bibinfo{author}{S.~Q. Fan}, \bibinfo{author}{B.~S.
  Hu}, \bibinfo{author}{J.~G. Li}, \bibinfo{author}{S.~Zhang},
  \bibinfo{author}{S.~M. Wang}, \bibinfo{author}{Z.~H. Sun},
  \bibinfo{author}{Y.~Z. Ma}, \bibinfo{author}{F.~R. Xu},
\newblock \bibinfo{title}{Deformed in-medium similarity renormalization group},
\newblock \bibinfo{journal}{Phys. Rev. C} \bibinfo{volume}{105}
  (\bibinfo{year}{2022}) \bibinfo{pages}{L061303}. \URLprefix
  \url{https://link.aps.org/doi/10.1103/PhysRevC.105.L061303}.
  \DOIprefix\doi{10.1103/PhysRevC.105.L061303}.
\bibitem[{Hu et~al.(2022)Hu, Jiang, Miyagi, Sun, Ekstr{\"o}m, Forss{\'e}n,
  Hagen, Holt, Papenbrock, Stroberg, and Vernon}]{B.S.Hu_Nat.Phys.12.186(2016)}
\bibinfo{author}{B.~S. Hu}, \bibinfo{author}{W.~G. Jiang},
  \bibinfo{author}{T.~Miyagi}, \bibinfo{author}{Z.~H. Sun},
  \bibinfo{author}{A.~Ekstr{\"o}m}, \bibinfo{author}{C.~Forss{\'e}n},
  \bibinfo{author}{G.~Hagen}, \bibinfo{author}{J.~D. Holt},
  \bibinfo{author}{T.~Papenbrock}, \bibinfo{author}{S.~R. Stroberg},
  \bibinfo{author}{I.~Vernon},
\newblock \bibinfo{title}{Ab initio predictions link the neutron skin of
  208{Pb} to nuclear forces},
\newblock \bibinfo{journal}{Nat. Phys.} \bibinfo{volume}{18}
  (\bibinfo{year}{2022}) \bibinfo{pages}{1196}. \URLprefix
  \url{https://doi.org/10.1038/s41567-022-01715-8}.
  \DOIprefix\doi{10.1038/s41567-022-01715-8}.
\bibitem[{Li et~al.(2019)Li, Michel, Hu, Zuo, and Xu}]{PhysRevC.100.054313}
\bibinfo{author}{J.~G. Li}, \bibinfo{author}{N.~Michel}, \bibinfo{author}{B.~S.
  Hu}, \bibinfo{author}{W.~Zuo}, \bibinfo{author}{F.~R. Xu},
\newblock \bibinfo{title}{Ab initio no-core gamow shell-model calculations of
  multineutron systems},
\newblock \bibinfo{journal}{Phys. Rev. C} \bibinfo{volume}{100}
  (\bibinfo{year}{2019}) \bibinfo{pages}{054313}. \URLprefix
  \url{https://link.aps.org/doi/10.1103/PhysRevC.100.054313}.
  \DOIprefix\doi{10.1103/PhysRevC.100.054313}.
\bibitem[{Tichai et~al.(2024)Tichai, Kap{\'a}s, Miyagi, Werner, Legeza,
  Schwenk, and Zarand}]{tichai2024}
\bibinfo{author}{A.~Tichai}, \bibinfo{author}{K.~Kap{\'a}s},
  \bibinfo{author}{T.~Miyagi}, \bibinfo{author}{M.~Werner},
  \bibinfo{author}{{\"O}.~Legeza}, \bibinfo{author}{A.~Schwenk},
  \bibinfo{author}{G.~Zarand},
\newblock \bibinfo{title}{Spectroscopy of ${N}=50$ isotones with the
  valence-space density matrix renormalization group},
\newblock \bibinfo{journal}{Phys. Lett. B} \bibinfo{volume}{855}
  (\bibinfo{year}{2024}) \bibinfo{pages}{138841}. \URLprefix
  \url{https://www.sciencedirect.com/science/article/pii/S037026932400399X}.
  \DOIprefix\doi{10.1016/j.physletb.2024.138841}.
\bibitem[{Tsukiyama et~al.(2012)Tsukiyama, Bogner, and
  Schwenk}]{PhysRevC.85.061304}
\bibinfo{author}{K.~Tsukiyama}, \bibinfo{author}{S.~K. Bogner},
  \bibinfo{author}{A.~Schwenk},
\newblock \bibinfo{title}{In-medium similarity renormalization group for
  open-shell nuclei},
\newblock \bibinfo{journal}{Phys. Rev. C} \bibinfo{volume}{85}
  (\bibinfo{year}{2012}) \bibinfo{pages}{061304}. \URLprefix
  \url{https://link.aps.org/doi/10.1103/PhysRevC.85.061304}.
  \DOIprefix\doi{10.1103/PhysRevC.85.061304}.
\bibitem[{Hergert et~al.(2016)Hergert, Bogner, Morris, Schwenk, and
  Tsukiyama}]{Phys.Rep.621.165}
\bibinfo{author}{H.~Hergert}, \bibinfo{author}{S.~Bogner},
  \bibinfo{author}{T.~Morris}, \bibinfo{author}{A.~Schwenk},
  \bibinfo{author}{K.~Tsukiyama},
\newblock \bibinfo{title}{The in-medium similarity renormalization group: A
  novel ab initio method for nuclei},
\newblock \bibinfo{journal}{Phys. Rep.} \bibinfo{volume}{621}
  (\bibinfo{year}{2016}) \bibinfo{pages}{165--222}. \URLprefix
  \url{https://www.sciencedirect.com/science/article/pii/S0370157315005414}.
  \DOIprefix\doi{https://doi.org/10.1016/j.physrep.2015.12.007}.
\bibitem[{Stroberg et~al.(2017)Stroberg, Calci, Hergert, Holt, Bogner, Roth,
  and Schwenk}]{PhysRevLett.118.032502}
\bibinfo{author}{S.~R. Stroberg}, \bibinfo{author}{A.~Calci},
  \bibinfo{author}{H.~Hergert}, \bibinfo{author}{J.~D. Holt},
  \bibinfo{author}{S.~K. Bogner}, \bibinfo{author}{R.~Roth},
  \bibinfo{author}{A.~Schwenk},
\newblock \bibinfo{title}{Nucleus-dependent valence-space approach to nuclear
  structure},
\newblock \bibinfo{journal}{Phys. Rev. Lett.} \bibinfo{volume}{118}
  (\bibinfo{year}{2017}) \bibinfo{pages}{032502}. \URLprefix
  \url{https://link.aps.org/doi/10.1103/PhysRevLett.118.032502}.
  \DOIprefix\doi{10.1103/PhysRevLett.118.032502}.
\bibitem[{Li et~al.(2023)Li, Yuan, Li, Xie, Zhang, Zhang, Xu, Michel, Xu, and
  Zuo}]{PhysRevC.107.014302}
\bibinfo{author}{H.~H. Li}, \bibinfo{author}{Q.~Yuan}, \bibinfo{author}{J.~G.
  Li}, \bibinfo{author}{M.~R. Xie}, \bibinfo{author}{S.~Zhang},
  \bibinfo{author}{Y.~H. Zhang}, \bibinfo{author}{X.~X. Xu},
  \bibinfo{author}{N.~Michel}, \bibinfo{author}{F.~R. Xu},
  \bibinfo{author}{W.~Zuo},
\newblock \bibinfo{title}{Investigation of isospin-symmetry breaking in mirror
  energy difference and nuclear mass with ab initio calculations},
\newblock \bibinfo{journal}{Phys. Rev. C} \bibinfo{volume}{107}
  (\bibinfo{year}{2023}) \bibinfo{pages}{014302}. \URLprefix
  \url{https://link.aps.org/doi/10.1103/PhysRevC.107.014302}.
  \DOIprefix\doi{10.1103/PhysRevC.107.014302}.
\bibitem[{Yuan et~al.(2024{\natexlab{a}})Yuan, Li, and Li}]{YUAN2024138331}
\bibinfo{author}{Q.~Yuan}, \bibinfo{author}{J.~G. Li}, \bibinfo{author}{H.~H.
  Li},
\newblock \bibinfo{title}{Ab initio calculations for well deformed nuclei:
  40{Mg} and 42{Si}},
\newblock \bibinfo{journal}{Phys. Lett. B} \bibinfo{volume}{848}
  (\bibinfo{year}{2024}{\natexlab{a}}) \bibinfo{pages}{138331}. \URLprefix
  \url{https://www.sciencedirect.com/science/article/pii/S0370269323006652}.
  \DOIprefix\doi{https://doi.org/10.1016/j.physletb.2023.138331}.
\bibitem[{Yuan et~al.(2024{\natexlab{b}})Yuan, Li, and
  Zuo}]{PhysRevC.109.L041301}
\bibinfo{author}{Q.~Yuan}, \bibinfo{author}{J.~G. Li},
  \bibinfo{author}{W.~Zuo},
\newblock \bibinfo{title}{Ab initio calculations for configuration-coexisting
  states in $^{45}\mathrm{S}$: An extension from $^{43}\mathrm{S}$},
\newblock \bibinfo{journal}{Phys. Rev. C} \bibinfo{volume}{109}
  (\bibinfo{year}{2024}{\natexlab{b}}) \bibinfo{pages}{L041301}. \URLprefix
  \url{https://link.aps.org/doi/10.1103/PhysRevC.109.L041301}.
  \DOIprefix\doi{10.1103/PhysRevC.109.L041301}.
\bibitem[{Hebeler et~al.(2011)Hebeler, Bogner, Furnstahl, Nogga, and
  Schwenk}]{PhysRevC.83.031301}
\bibinfo{author}{K.~Hebeler}, \bibinfo{author}{S.~K. Bogner},
  \bibinfo{author}{R.~J. Furnstahl}, \bibinfo{author}{A.~Nogga},
  \bibinfo{author}{A.~Schwenk},
\newblock \bibinfo{title}{Improved nuclear matter calculations from chiral
  low-momentum interactions},
\newblock \bibinfo{journal}{Phys. Rev. C} \bibinfo{volume}{83}
  (\bibinfo{year}{2011}) \bibinfo{pages}{031301(R)}. \URLprefix
  \url{https://link.aps.org/doi/10.1103/PhysRevC.83.031301}.
  \DOIprefix\doi{10.1103/PhysRevC.83.031301}.
\bibitem[{Jiang et~al.(2020)Jiang, Ekstr\"om, Forss\'en, Hagen, Jansen, and
  Papenbrock}]{PhysRevC.102.054301}
\bibinfo{author}{W.~G. Jiang}, \bibinfo{author}{A.~Ekstr\"om},
  \bibinfo{author}{C.~Forss\'en}, \bibinfo{author}{G.~Hagen},
  \bibinfo{author}{G.~R. Jansen}, \bibinfo{author}{T.~Papenbrock},
\newblock \bibinfo{title}{Accurate bulk properties of nuclei from ${A}=2$ to
  $\ensuremath{\infty}$ from potentials with $\mathrm{\ensuremath{\Delta}}$
  isobars},
\newblock \bibinfo{journal}{Phys. Rev. C} \bibinfo{volume}{102}
  (\bibinfo{year}{2020}) \bibinfo{pages}{054301}. \URLprefix
  \url{https://link.aps.org/doi/10.1103/PhysRevC.102.054301}.
  \DOIprefix\doi{10.1103/PhysRevC.102.054301}.
\bibitem[{Hagen et~al.(2016)Hagen, Jansen, and Papenbrock}]{hagen2016b}
\bibinfo{author}{G.~Hagen}, \bibinfo{author}{G.~R. Jansen},
  \bibinfo{author}{T.~Papenbrock},
\newblock \bibinfo{title}{Structure of $^{78}\mathrm{Ni}$ from first-principles
  computations},
\newblock \bibinfo{journal}{Phys. Rev. Lett.} \bibinfo{volume}{117}
  (\bibinfo{year}{2016}) \bibinfo{pages}{172501}. \URLprefix
  \url{https://link.aps.org/doi/10.1103/PhysRevLett.117.172501}.
  \DOIprefix\doi{10.1103/PhysRevLett.117.172501}.
\bibitem[{Simonis et~al.(2017)Simonis, Stroberg, Hebeler, Holt, and
  Schwenk}]{simonis2017}
\bibinfo{author}{J.~Simonis}, \bibinfo{author}{S.~R. Stroberg},
  \bibinfo{author}{K.~Hebeler}, \bibinfo{author}{J.~D. Holt},
  \bibinfo{author}{A.~Schwenk},
\newblock \bibinfo{title}{Saturation with chiral interactions and consequences
  for finite nuclei},
\newblock \bibinfo{journal}{Phys. Rev. C} \bibinfo{volume}{96}
  (\bibinfo{year}{2017}) \bibinfo{pages}{014303}. \URLprefix
  \url{https://link.aps.org/doi/10.1103/PhysRevC.96.014303}.
  \DOIprefix\doi{10.1103/PhysRevC.96.014303}.
\bibitem[{Stroberg et~al.(2021)Stroberg, Holt, Schwenk, and
  Simonis}]{PhysRevLett.126.022501}
\bibinfo{author}{S.~R. Stroberg}, \bibinfo{author}{J.~D. Holt},
  \bibinfo{author}{A.~Schwenk}, \bibinfo{author}{J.~Simonis},
\newblock \bibinfo{title}{Ab initio limits of atomic nuclei},
\newblock \bibinfo{journal}{Phys. Rev. Lett.} \bibinfo{volume}{126}
  (\bibinfo{year}{2021}) \bibinfo{pages}{022501}. \URLprefix
  \url{https://link.aps.org/doi/10.1103/PhysRevLett.126.022501}.
  \DOIprefix\doi{10.1103/PhysRevLett.126.022501}.
\bibitem[{Miyagi et~al.(2022)Miyagi, Stroberg, Navr\'atil, Hebeler, and
  Holt}]{PhysRevC.105.014302}
\bibinfo{author}{T.~Miyagi}, \bibinfo{author}{S.~R. Stroberg},
  \bibinfo{author}{P.~Navr\'atil}, \bibinfo{author}{K.~Hebeler},
  \bibinfo{author}{J.~D. Holt},
\newblock \bibinfo{title}{Converged ab initio calculations of heavy nuclei},
\newblock \bibinfo{journal}{Phys. Rev. C} \bibinfo{volume}{105}
  (\bibinfo{year}{2022}) \bibinfo{pages}{014302}. \URLprefix
  \url{https://link.aps.org/doi/10.1103/PhysRevC.105.014302}.
  \DOIprefix\doi{10.1103/PhysRevC.105.014302}.
\bibitem[{Hebeler et~al.(2023)Hebeler, Durant, Hoppe, Heinz, Schwenk, Simonis,
  and Tichai}]{hebeler2023}
\bibinfo{author}{K.~Hebeler}, \bibinfo{author}{V.~Durant},
  \bibinfo{author}{J.~Hoppe}, \bibinfo{author}{M.~Heinz},
  \bibinfo{author}{A.~Schwenk}, \bibinfo{author}{J.~Simonis},
  \bibinfo{author}{A.~Tichai},
\newblock \bibinfo{title}{Normal ordering of three-nucleon interactions for ab
  initio calculations of heavy nuclei},
\newblock \bibinfo{journal}{Phys. Rev. C} \bibinfo{volume}{107}
  (\bibinfo{year}{2023}) \bibinfo{pages}{024310}. \URLprefix
  \url{https://link.aps.org/doi/10.1103/PhysRevC.107.024310}.
  \DOIprefix\doi{10.1103/PhysRevC.107.024310}.
\bibitem[{Hagen et~al.(2007)Hagen, Papenbrock, Dean, Schwenk, Nogga, W\l{}och,
  and Piecuch}]{hagen2007a}
\bibinfo{author}{G.~Hagen}, \bibinfo{author}{T.~Papenbrock},
  \bibinfo{author}{D.~J. Dean}, \bibinfo{author}{A.~Schwenk},
  \bibinfo{author}{A.~Nogga}, \bibinfo{author}{M.~W\l{}och},
  \bibinfo{author}{P.~Piecuch},
\newblock \bibinfo{title}{{Coupled-cluster theory for three-body
  Hamiltonians}},
\newblock \bibinfo{journal}{Phys. Rev. C} \bibinfo{volume}{76}
  (\bibinfo{year}{2007}) \bibinfo{pages}{034302}. \URLprefix
  \url{http://link.aps.org/doi/10.1103/PhysRevC.76.034302}.
  \DOIprefix\doi{10.1103/PhysRevC.76.034302}.
\bibitem[{Roth et~al.(2012)Roth, Binder, Vobig, Calci, Langhammer, and
  Navr\'atil}]{PhysRevLett.109.052501}
\bibinfo{author}{R.~Roth}, \bibinfo{author}{S.~Binder},
  \bibinfo{author}{K.~Vobig}, \bibinfo{author}{A.~Calci},
  \bibinfo{author}{J.~Langhammer}, \bibinfo{author}{P.~Navr\'atil},
\newblock \bibinfo{title}{Medium-mass nuclei with normal-ordered chiral
  ${NN}\mathbf{+}{3N}$ interactions},
\newblock \bibinfo{journal}{Phys. Rev. Lett.} \bibinfo{volume}{109}
  (\bibinfo{year}{2012}) \bibinfo{pages}{052501}. \URLprefix
  \url{https://link.aps.org/doi/10.1103/PhysRevLett.109.052501}.
  \DOIprefix\doi{10.1103/PhysRevLett.109.052501}.
\bibitem[{Morris et~al.(2015)Morris, Parzuchowski, and
  Bogner}]{PhysRevC.92.034331}
\bibinfo{author}{T.~D. Morris}, \bibinfo{author}{N.~M. Parzuchowski},
  \bibinfo{author}{S.~K. Bogner},
\newblock \bibinfo{title}{Magnus expansion and in-medium similarity
  renormalization group},
\newblock \bibinfo{journal}{Phys. Rev. C} \bibinfo{volume}{92}
  (\bibinfo{year}{2015}) \bibinfo{pages}{034331}. \URLprefix
  \url{https://link.aps.org/doi/10.1103/PhysRevC.92.034331}.
  \DOIprefix\doi{10.1103/PhysRevC.92.034331}.
\bibitem[{Shimizu et~al.(2019)Shimizu, Mizusaki, Utsuno, and
  Tsunoda}]{Phys.Commun.244.372}
\bibinfo{author}{N.~Shimizu}, \bibinfo{author}{T.~Mizusaki},
  \bibinfo{author}{Y.~Utsuno}, \bibinfo{author}{Y.~Tsunoda},
\newblock \bibinfo{title}{Thick-restart block lanczos method for large-scale
  shell-model calculations},
\newblock \bibinfo{journal}{Comput. Phys. Commun.} \bibinfo{volume}{244}
  (\bibinfo{year}{2019}) \bibinfo{pages}{372--384}. \URLprefix
  \url{https://www.sciencedirect.com/science/article/pii/S0010465519301985}.
  \DOIprefix\doi{https://doi.org/10.1016/j.cpc.2019.06.011}.
\bibitem[{nnd(2024)}]{nndc}
\bibinfo{howpublished}{Data extracted using the {NNDC} On-Line Data Service,
  http://www.nndc.bnl.gov/}, \bibinfo{year}{2024}. \URLprefix
  \url{http://www.nndc.bnl.gov/}.
\bibitem[{Park et~al.(2017)Park, Kr\"ucken, Lubos, Gernh\"auser, Lewitowicz,
  Nishimura, Ahn, Baba, Blank, Blazhev, Boutachkov, Browne, \ifmmode
  \check{C}\else \v{C}\fi{}elikovi\ifmmode~\acute{c}\else \'{c}\fi{},
  de~France, Doornenbal, Faestermann, Fang, Fukuda, Giovinazzo, Goel, G\'orska,
  Grawe, Ilieva, Inabe, Isobe, Jungclaus, Kameda, Kim, Kim, Kojouharov, Kubo,
  Kurz, Lorusso, Moschner, Murai, Nishizuka, Patel, Rajabali, Rice, Sakurai,
  Schaffner, Shimizu, Sinclair, S\"oderstr\"om, Steiger, Sumikama, Suzuki,
  Takeda, Wang, Watanabe, Wu, and Xu}]{PhysRevC.96.044311}
\bibinfo{author}{J.~Park}, \bibinfo{author}{R.~Kr\"ucken},
  \bibinfo{author}{D.~Lubos}, \bibinfo{author}{R.~Gernh\"auser},
  \bibinfo{author}{M.~Lewitowicz}, \bibinfo{author}{S.~Nishimura},
  \bibinfo{author}{D.~S. Ahn}, \bibinfo{author}{H.~Baba},
  \bibinfo{author}{B.~Blank}, \bibinfo{author}{A.~Blazhev},
  \bibinfo{author}{P.~Boutachkov}, \bibinfo{author}{F.~Browne},
  \bibinfo{author}{I.~\ifmmode \check{C}\else
  \v{C}\fi{}elikovi\ifmmode~\acute{c}\else \'{c}\fi{}},
  \bibinfo{author}{G.~de~France}, \bibinfo{author}{P.~Doornenbal},
  \bibinfo{author}{T.~Faestermann}, \bibinfo{author}{Y.~Fang},
  \bibinfo{author}{N.~Fukuda}, \bibinfo{author}{J.~Giovinazzo},
  \bibinfo{author}{N.~Goel}, \bibinfo{author}{M.~G\'orska},
  \bibinfo{author}{H.~Grawe}, \bibinfo{author}{S.~Ilieva},
  \bibinfo{author}{N.~Inabe}, \bibinfo{author}{T.~Isobe},
  \bibinfo{author}{A.~Jungclaus}, \bibinfo{author}{D.~Kameda},
  \bibinfo{author}{G.~D. Kim}, \bibinfo{author}{Y.-K. Kim},
  \bibinfo{author}{I.~Kojouharov}, \bibinfo{author}{T.~Kubo},
  \bibinfo{author}{N.~Kurz}, \bibinfo{author}{G.~Lorusso},
  \bibinfo{author}{K.~Moschner}, \bibinfo{author}{D.~Murai},
  \bibinfo{author}{I.~Nishizuka}, \bibinfo{author}{Z.~Patel},
  \bibinfo{author}{M.~M. Rajabali}, \bibinfo{author}{S.~Rice},
  \bibinfo{author}{H.~Sakurai}, \bibinfo{author}{H.~Schaffner},
  \bibinfo{author}{Y.~Shimizu}, \bibinfo{author}{L.~Sinclair},
  \bibinfo{author}{P.-A. S\"oderstr\"om}, \bibinfo{author}{K.~Steiger},
  \bibinfo{author}{T.~Sumikama}, \bibinfo{author}{H.~Suzuki},
  \bibinfo{author}{H.~Takeda}, \bibinfo{author}{Z.~Wang},
  \bibinfo{author}{H.~Watanabe}, \bibinfo{author}{J.~Wu},
  \bibinfo{author}{Z.~Y. Xu},
\newblock \bibinfo{title}{Properties of $\ensuremath{\gamma}$-decaying isomers
  and isomeric ratios in the $^{100}\mathrm{Sn}$ region},
\newblock \bibinfo{journal}{Phys. Rev. C} \bibinfo{volume}{96}
  (\bibinfo{year}{2017}) \bibinfo{pages}{044311}. \URLprefix
  \url{https://link.aps.org/doi/10.1103/PhysRevC.96.044311}.
  \DOIprefix\doi{10.1103/PhysRevC.96.044311}.
\bibitem[{Park et~al.(2021)Park, Kr\"ucken, Lubos, Gernh\"auser, Lewitowicz,
  Nishimura, Ahn, Baba, Blank, Blazhev, Boutachkov, Browne, \ifmmode
  \check{C}\else \v{C}\fi{}elikovi\ifmmode~\acute{c}\else \'{c}\fi{},
  de~France, Doornenbal, Faestermann, Fang, Fukuda, Giovinazzo, Goel, G\'orska,
  Grawe, Ilieva, Inabe, Isobe, Jungclaus, Kameda, Kim, Kim, Kojouharov, Kubo,
  Kurz, Lorusso, Moschner, Murai, Nishizuka, Patel, Rajabali, Rice, Sakurai,
  Schaffner, Shimizu, Sinclair, S\"oderstr\"om, Steiger, Sumikama, Suzuki,
  Takeda, Wang, Watanabe, Wu, and Xu}]{PhysRevC.103.049901}
\bibinfo{author}{J.~Park}, \bibinfo{author}{R.~Kr\"ucken},
  \bibinfo{author}{D.~Lubos}, \bibinfo{author}{R.~Gernh\"auser},
  \bibinfo{author}{M.~Lewitowicz}, \bibinfo{author}{S.~Nishimura},
  \bibinfo{author}{D.~S. Ahn}, \bibinfo{author}{H.~Baba},
  \bibinfo{author}{B.~Blank}, \bibinfo{author}{A.~Blazhev},
  \bibinfo{author}{P.~Boutachkov}, \bibinfo{author}{F.~Browne},
  \bibinfo{author}{I.~\ifmmode \check{C}\else
  \v{C}\fi{}elikovi\ifmmode~\acute{c}\else \'{c}\fi{}},
  \bibinfo{author}{G.~de~France}, \bibinfo{author}{P.~Doornenbal},
  \bibinfo{author}{T.~Faestermann}, \bibinfo{author}{Y.~Fang},
  \bibinfo{author}{N.~Fukuda}, \bibinfo{author}{J.~Giovinazzo},
  \bibinfo{author}{N.~Goel}, \bibinfo{author}{M.~G\'orska},
  \bibinfo{author}{H.~Grawe}, \bibinfo{author}{S.~Ilieva},
  \bibinfo{author}{N.~Inabe}, \bibinfo{author}{T.~Isobe},
  \bibinfo{author}{A.~Jungclaus}, \bibinfo{author}{D.~Kameda},
  \bibinfo{author}{G.~D. Kim}, \bibinfo{author}{Y.-K. Kim},
  \bibinfo{author}{I.~Kojouharov}, \bibinfo{author}{T.~Kubo},
  \bibinfo{author}{N.~Kurz}, \bibinfo{author}{G.~Lorusso},
  \bibinfo{author}{K.~Moschner}, \bibinfo{author}{D.~Murai},
  \bibinfo{author}{I.~Nishizuka}, \bibinfo{author}{Z.~Patel},
  \bibinfo{author}{M.~M. Rajabali}, \bibinfo{author}{S.~Rice},
  \bibinfo{author}{H.~Sakurai}, \bibinfo{author}{H.~Schaffner},
  \bibinfo{author}{Y.~Shimizu}, \bibinfo{author}{L.~Sinclair},
  \bibinfo{author}{P.-A. S\"oderstr\"om}, \bibinfo{author}{K.~Steiger},
  \bibinfo{author}{T.~Sumikama}, \bibinfo{author}{H.~Suzuki},
  \bibinfo{author}{H.~Takeda}, \bibinfo{author}{Z.~Wang},
  \bibinfo{author}{H.~Watanabe}, \bibinfo{author}{J.~Wu},
  \bibinfo{author}{Z.~Y. Xu},
\newblock \bibinfo{title}{Erratum: Properties of $\ensuremath{\gamma}$-decaying
  isomers and isomeric ratios in the $^{100}\mathrm{Sn}$ region [phys. rev. c
  96, 044311 (2017)]},
\newblock \bibinfo{journal}{Phys. Rev. C} \bibinfo{volume}{103}
  (\bibinfo{year}{2021}) \bibinfo{pages}{049901}. \URLprefix
  \url{https://link.aps.org/doi/10.1103/PhysRevC.103.049901}.
  \DOIprefix\doi{10.1103/PhysRevC.103.049901}.
\bibitem[{Ley et~al.(2023)Ley, Knafla, Jolie, Esmaylzadeh, Harter, Blazhev,
  Fransen, Pfeil, R\'egis, and Van~Isacker}]{PhysRevC.108.064313}
\bibinfo{author}{M.~Ley}, \bibinfo{author}{L.~Knafla},
  \bibinfo{author}{J.~Jolie}, \bibinfo{author}{A.~Esmaylzadeh},
  \bibinfo{author}{A.~Harter}, \bibinfo{author}{A.~Blazhev},
  \bibinfo{author}{C.~Fransen}, \bibinfo{author}{A.~Pfeil},
  \bibinfo{author}{J.-M. R\'egis}, \bibinfo{author}{P.~Van~Isacker},
\newblock \bibinfo{title}{Lifetime measurements in $^{92}\mathrm{Mo}$:
  Investigation of seniority conservation in the ${N}=50$ isotones},
\newblock \bibinfo{journal}{Phys. Rev. C} \bibinfo{volume}{108}
  (\bibinfo{year}{2023}) \bibinfo{pages}{064313}. \URLprefix
  \url{https://link.aps.org/doi/10.1103/PhysRevC.108.064313}.
  \DOIprefix\doi{10.1103/PhysRevC.108.064313}.
\bibitem[{Sun et~al.(2024)Sun, Ekström, Forssén, Hagen, Jansen, and
  Papenbrock}]{sun2024}
\bibinfo{author}{Z.~H. Sun}, \bibinfo{author}{A.~Ekström},
  \bibinfo{author}{C.~Forssén}, \bibinfo{author}{G.~Hagen},
  \bibinfo{author}{G.~R. Jansen}, \bibinfo{author}{T.~Papenbrock},
\newblock \bibinfo{title}{Multiscale physics of atomic nuclei from first
  principles},
\newblock \bibinfo{journal}{arXiv:2404.00058 [nucl-th]}
  (\bibinfo{year}{2024}). \URLprefix \url{https://arxiv.org/abs/2404.00058}.
  \href{http://arxiv.org/abs/2404.00058}{{\tt arXiv:2404.00058}}.
\bibitem[{Henderson et~al.(2018)Henderson, Hackman, Ruotsalainenand
  et~al.}]{Hend18E2}
\bibinfo{author}{J.~Henderson}, \bibinfo{author}{G.~Hackman},
  \bibinfo{author}{P.~Ruotsalainenand}, et~al.,
\newblock \bibinfo{title}{{Testing microscopically derived descriptions of
  nuclear collectivity: Coulomb excitation of $^{22}\mathrm{Mg}$}},
\newblock \bibinfo{journal}{Phys. Lett. B} \bibinfo{volume}{782}
  (\bibinfo{year}{2018}) \bibinfo{pages}{468--473}. \URLprefix
  \url{https://doi.org/10.1016/j.physletb.2018.05.064}.
  \DOIprefix\doi{10.1016/j.physletb.2018.05.064}.
\bibitem[{Bai et~al.(2022)Bai, Koszor{\'{u}}s, Hu et~al.}]{BAI2022137064}
\bibinfo{author}{S.~Bai}, \bibinfo{author}{{\'{A}}.~Koszor{\'{u}}s},
  \bibinfo{author}{B.~Hu}, et~al.,
\newblock \bibinfo{title}{{Electromagnetic moments of scandium isotopes and N =
  28 isotones in the distinctive 0f7/2 orbit}},
\newblock \bibinfo{journal}{Phys. Lett. B} \bibinfo{volume}{829}
  (\bibinfo{year}{2022}) \bibinfo{pages}{137064}. \URLprefix
  \url{https://www.sciencedirect.com/science/article/pii/S0370269322001988
  https://linkinghub.elsevier.com/retrieve/pii/S0370269322001988}.
  \DOIprefix\doi{10.1016/j.physletb.2022.137064}.
\bibitem[{Hu et~al.(2024)Hu, Sun, Hagen, and Papenbrock}]{hu2024}
\bibinfo{author}{B.~S. Hu}, \bibinfo{author}{Z.~H. Sun},
  \bibinfo{author}{G.~Hagen}, \bibinfo{author}{T.~Papenbrock},
\newblock \bibinfo{title}{Ab initio computations of strongly deformed nuclei
  near $^{80}\mathrm{Zr}$},
\newblock \bibinfo{journal}{Phys. Rev. C} \bibinfo{volume}{110}
  (\bibinfo{year}{2024}) \bibinfo{pages}{L011302}. \URLprefix
  \url{https://link.aps.org/doi/10.1103/PhysRevC.110.L011302}.
  \DOIprefix\doi{10.1103/PhysRevC.110.L011302}.
\bibitem[{Blazhev et~al.(2004)Blazhev, G\'orska, Grawe, Nyberg, Palacz,
  Caurier, Dorvaux, Gadea, Nowacki, Andreoiu, de~Angelis, Balabanski, Beck,
  Cederwall, Curien, D\"oring, Ekman, Fahlander, Lagergren, Ljungvall,
  Moszy\'nski, Norlin, Plettner, Rudolph, Sohler, Spohr, Thelen, Weiszflog,
  Wisell, Woli\'nska, and Wolski}]{PhysRevC.69.064304}
\bibinfo{author}{A.~Blazhev}, \bibinfo{author}{M.~G\'orska},
  \bibinfo{author}{H.~Grawe}, \bibinfo{author}{J.~Nyberg},
  \bibinfo{author}{M.~Palacz}, \bibinfo{author}{E.~Caurier},
  \bibinfo{author}{O.~Dorvaux}, \bibinfo{author}{A.~Gadea},
  \bibinfo{author}{F.~Nowacki}, \bibinfo{author}{C.~Andreoiu},
  \bibinfo{author}{G.~de~Angelis}, \bibinfo{author}{D.~Balabanski},
  \bibinfo{author}{C.~Beck}, \bibinfo{author}{B.~Cederwall},
  \bibinfo{author}{D.~Curien}, \bibinfo{author}{J.~D\"oring},
  \bibinfo{author}{J.~Ekman}, \bibinfo{author}{C.~Fahlander},
  \bibinfo{author}{K.~Lagergren}, \bibinfo{author}{J.~Ljungvall},
  \bibinfo{author}{M.~Moszy\'nski}, \bibinfo{author}{L.-O. Norlin},
  \bibinfo{author}{C.~Plettner}, \bibinfo{author}{D.~Rudolph},
  \bibinfo{author}{D.~Sohler}, \bibinfo{author}{K.~M. Spohr},
  \bibinfo{author}{O.~Thelen}, \bibinfo{author}{M.~Weiszflog},
  \bibinfo{author}{M.~Wisell}, \bibinfo{author}{M.~Woli\'nska},
  \bibinfo{author}{W.~Wolski},
\newblock \bibinfo{title}{Observation of a core-excited ${E4}$ isomer in
  $^{98}\mathrm{Cd}$},
\newblock \bibinfo{journal}{Phys. Rev. C} \bibinfo{volume}{69}
  (\bibinfo{year}{2004}) \bibinfo{pages}{064304}. \URLprefix
  \url{https://link.aps.org/doi/10.1103/PhysRevC.69.064304}.
  \DOIprefix\doi{10.1103/PhysRevC.69.064304}.
\bibitem[{Jungclaus et~al.(1998)Jungclaus, Kast, Lieb, Teich, Weiszflog,
  Härtlein, Ender, Köck, Schwalm, Reif, Peusquens, Dewald, Eberth, Thomas,
  G\'orska, and Grawe}]{JUNGCLAUS1998346}
\bibinfo{author}{A.~Jungclaus}, \bibinfo{author}{D.~Kast},
  \bibinfo{author}{K.~P. Lieb}, \bibinfo{author}{C.~Teich},
  \bibinfo{author}{M.~Weiszflog}, \bibinfo{author}{T.~Härtlein},
  \bibinfo{author}{C.~Ender}, \bibinfo{author}{F.~Köck},
  \bibinfo{author}{D.~Schwalm}, \bibinfo{author}{J.~Reif},
  \bibinfo{author}{R.~Peusquens}, \bibinfo{author}{A.~Dewald},
  \bibinfo{author}{J.~Eberth}, \bibinfo{author}{H.~G. Thomas},
  \bibinfo{author}{M.~G\'orska}, \bibinfo{author}{H.~Grawe},
\newblock \bibinfo{title}{Picosecond lifetime measurement of neutron
  core-excited states in the {N} = 50 nucleus 95rh},
\newblock \bibinfo{journal}{Nucl. Phys. A} \bibinfo{volume}{637}
  (\bibinfo{year}{1998}) \bibinfo{pages}{346--364}. \URLprefix
  \url{https://www.sciencedirect.com/science/article/pii/S0375947498002322}.
  \DOIprefix\doi{https://doi.org/10.1016/S0375-9474(98)00232-2}.
\bibitem[{Miyagi(2023)}]{miyagi2023}
\bibinfo{author}{T.~Miyagi},
\newblock \bibinfo{title}{Nuhamil: A numerical code to generate nuclear two-
  and three-body matrix elements from chiral effective field theory},
\newblock \bibinfo{journal}{Eur. Phys. J. A} \bibinfo{volume}{59}
  (\bibinfo{year}{2023}) \bibinfo{pages}{150}. \URLprefix
  \url{https://doi.org/10.1140/epja/s10050-023-01039-y}.
  \DOIprefix\doi{10.1140/epja/s10050-023-01039-y}.
\bibitem[{Stroberg(2024)}]{Stro17imsrg++}
\bibinfo{author}{S.~R. Stroberg},
  \bibinfo{title}{https://github.com/ragnarstroberg/imsrg},
  \bibinfo{year}{2024}.
  \DOIprefix\doi{https://github.com/ragnarstroberg/imsrg}.

\end{thebibliography}

\end{document}